\documentclass[11pt]{article}

\usepackage[numbers,sort&compress]{natbib}
\usepackage{graphicx}
\usepackage{amsmath,graphicx,amssymb,subfigure}
\usepackage{amssymb}
\usepackage{amsmath}
\usepackage{enumerate}
\usepackage{amsfonts}
\usepackage[all]{xy}
\usepackage[section]{placeins}
\usepackage[colorlinks=true,linktocpage=true,linkcolor=blue,citecolor=blue]{hyperref}
\newcommand{\mt}[1]{\textrm{\tiny #1}}
\newcommand{\bea}{\begin{eqnarray}}
\newcommand{\eea}{\end{eqnarray}}
\newcommand{\cf}{{\cal F}}
\newcommand{\cb}{{\cal B}}
\newcommand{\ch}{{\cal H}}
\newcommand{\uh}{u_\mt{H}}
\newcommand{\ta}{\theta}

\begin{document}

\bibliographystyle{hieeetr}

\pagestyle{plain}
\setcounter{page}{1}

\begin{titlepage}

\begin{center}

\vskip 20mm

{\LARGE {\bf Drag force of Anisotropic plasma at finite $U(1)$ chemical potential}}

\vskip 1 cm

{\large {\bf Long Cheng$^1$ ,  Xian-Hui Ge$^1$, Shang-Yu Wu$^{2,3}$}}\\

\vskip .8cm

{\it $^1$Department of Physics, Shanghai University, Shanghai 200444,  China}\\
{\it $^2$Department of Electrophysics, Yau Shing Tung Center, National Chiao Tung University, Hsinchu 300, Taiwan.\\
$^3$National Center for Theoretical Science, Hsinchu, Taiwan.}

\medskip

\vskip 0.5cm

{\tt  physcheng@shu.edu.cn, \, gexh@shu.edu.cn, loganwu@gmail.com}

\vspace{5mm}
\vspace{5mm}

{\bf Abstract}\\

\end{center}
\noindent
We perform the calculation of drag force acting on a massive quark moving through an anisotropic ${\cal N}=4$ SU(N) Super Yang-Mills plasma in the presence of a $U(1)$ chemical potential. We present the numerical results for any  value of anisotropy and arbitrary direction of the quark velocity with respect to the direction of anisotropy. We find the effect of chemical potential or charge density will enhance the drag force for the our charged solution.

\end{titlepage}

\section{Introduction}
The experimental data in the Relativistic Heavy Ion Collider (RHIC) \cite{rhic,rhic2} show that the quark gluon plasma (QGP), as deconfined phase of QCD at
high temperature and high number density, is the strongly coupled fluid rather than the weakly coupled gas of quarks and gluons. Thus the perturbative QCD is no longer reliable
and we should explore the non-perturbative  methods of QCD.

The AdS/CFT correspondence \cite{duality,duality2,duality3} provides a powerful tool to investigate the strongly coupled system in condensed matter physics (for reviews, see\cite{Hartnoll,McGreevy}) like superconductors \cite{Hartnoll:2008vx, Wen:2013ufa,gw}, Lifshitz fixed point \cite{Kachru:2008yh, Sun:2013wpa, Sun:2013zga,fang}, Quantum chromodynamics \cite{review1,review2} and heavy ion collisions like photon production \cite{CaronHuot:2006te, Wu:2013qja}, elliptic flow \cite{Muller:2013ila}, drag force \cite{drag1, drag2}, jet quenching \cite{jet1, jet2}, Langevin coefficients \cite{gil} and anomalous transport \cite{Yee:2009vw, Kharzeev:2010gd, Pu:2014cwa, Pu:2014fva}. Using the AdS/CFT correspondence , we can study the strongly ${\cal N} = 4 $ Super-Yang-Mills plasma through considering IIB supergravity in $AdS_5\times S^5$. One significant result is the calculation of the ratio of shear viscosity to entropy density of QGP, this ratio is universal and equal to $1/4\pi$ \cite{dts}, which matches the experimental data very well. This indicates the QGP is a strongly coupled system and the AdS/CFT presents a useful method to investigate the properties of QGP at least at qualitative level.

It is well-known that the most important quantities of QGP are drag force and jet-quenching parameter.
In the context of AdS/CFT \cite{drag1,drag2}, the moving heavy quark in the thermal medium is dual to a probe open string with infinite mass, which is attached to the boundary of bulk space-time and stretch to the black hole horizon. So the dynamics of string can give us the effects of ${\cal N} = 4 $ Super-Yang-Mills plasma in which quark is moving.
The similar study for jet quenching for ${\cal N} = 4 $ SYM plasma  can be found in \cite{jet1,jet2}, where the jet-quenching parameter was obtained by calculating
the expectation value of a closed light-like Wilson loop in the dipole approximation.

In this paper, we will study the moving quark in the anisotropic QGP with chemical potential, since the QGP after creation in a short time is anisotropic and real experiments are done at finite baryon potential \footnote{But the model we consider is static and the origin of anisotropy in this model is different than the QGP anisotropy.}. Our motivations come from the facts that significant observation of the RHIC and the LHC experiments is that the plasma created is anisotropic and non-equilibrium during the period of time $\tau_{out}$ after the collision, i.e. it is locally anisotropic at the time $\tau_{out} <\tau<\tau_{iso}$, a configuration to be described by the hydrodynamics with the anisotropic energy-momentum tensor.

Another motivation comes from the fact that in condensed matter physics, some materials are anisotropic, with different properties in different directions. For example, for   high temperature cuprates, the crystal structure of such superconductors shows muti-layer of $CuO_2$ planes with superconductivity taking place between these layers.  The electric transport perpendicular to the $CuO_2$ plane is more difficult than the electric transport in the $CuO_2$ plane \cite{cuprates}. It is natural to investigate drag force in anisotropic system at finite chemical and compare the results with those isotropic cases.

From the point of view of holography, the anisotropic plasma with chemical potential is dual to a anisotropic charged black brane
\cite{ch1,ch2}, hence to compute  the drag force experienced by an infinitely massive quark, we should consider a string in anisotropic charged black brane. We will show the results analytically  and numerically respectively. The discussion for chargeless anisotropic plasma, shear viscosity-entropy density ratio and its energy loss in the framework of AdS/CFT can be found in \cite{ma1,ma2,rehban,ma3,gi,ma4, Wu:2013qja}.

This paper is organized as follows. In Section 2, we briefly describe the construction of the anisotropic charged black brane solution. In Section 3, we calculate the drag force acting on a massive quark moving through the plasma. Furthermore, in the small anisotropy and high temperature limit, we can perform the calculation of the drag force analytically. In Section 4, we show our numerical results for the prolate anisotropy. In Section 5, we conclude with a brief summary of our results.

Note: When this paper is in the preparation, the paper \cite{Chakraborty:2014kfa} appeared which has some overlap with this paper. The differences between our paper and \cite{Chakraborty:2014kfa}, are that we consider the drag force for  the large charge case in particular.

\section{Anisotropic charged black brane solution}
The five dimensional axion-dilaton-Maxwell-gravity bulk action reduced from type IIB supergravity is given by \cite{ch1,ch2}
\bea\label{5action}
S=\frac{1}{2\kappa^2}\bigg[\int d^5 x \sqrt{-g} \Big( R +12-\frac{1}{2} (\partial\phi)^2 - \frac{1}{2} e^{2\phi} (\partial \chi)^2- \frac{1}{4} F_{\mu\nu}F^{\mu\nu}\Big)-2\int d^4x \sqrt{-\gamma}K \bigg]\,,
\eea
where we have set the AdS radius $L=1$, and $\kappa^2=8\pi G=\frac{4\pi^2}{N^2_c}$.
The $AdS_5$ part of ten dimensional anisotropic IIB supergravity solution in Einstein frame is given by \cite{ch2},
\bea
&&\hskip -.35cm
ds^2 =  \frac{e^{-\frac{1}{2}\phi}}{u^2}\left( -\cf \cb\, dt^2+dx^2+dy^2+ \ch dz^2 +\frac{ du^2}{\cf}\right),\\
&& A=A_t(u) dt,~~~{\rm and~~~~ } \chi=az,
\,\,\,\,\,\,\label{ansatz1}
\eea
which was obtained by dimensional reduction on $S^5$ from the ten-dimensional metric.  So we have ignored the metric of the $S^5$ part that will play no role in the following discussions and our treatment is in agreement with ten dimensional framework. Note that the anisotropy is introduced through deforming the SYM theory by a $\theta$-parameter of the form $\theta\propto z$, which acts as an isotropy-breaking external source that forces the system into an anisotropic equilibrium state \cite{ma1}. The $\theta$-parameter is dual to the type IIB axion $\chi$ with the form $\chi = a z$ \footnote{The model and solutions obtained in the following are the direct generalization of \cite{ma1, ma2}.}.

\begin{figure}[htbp]
 \begin{minipage}{1\hsize}
\begin{center}
\includegraphics*[scale=0.50] {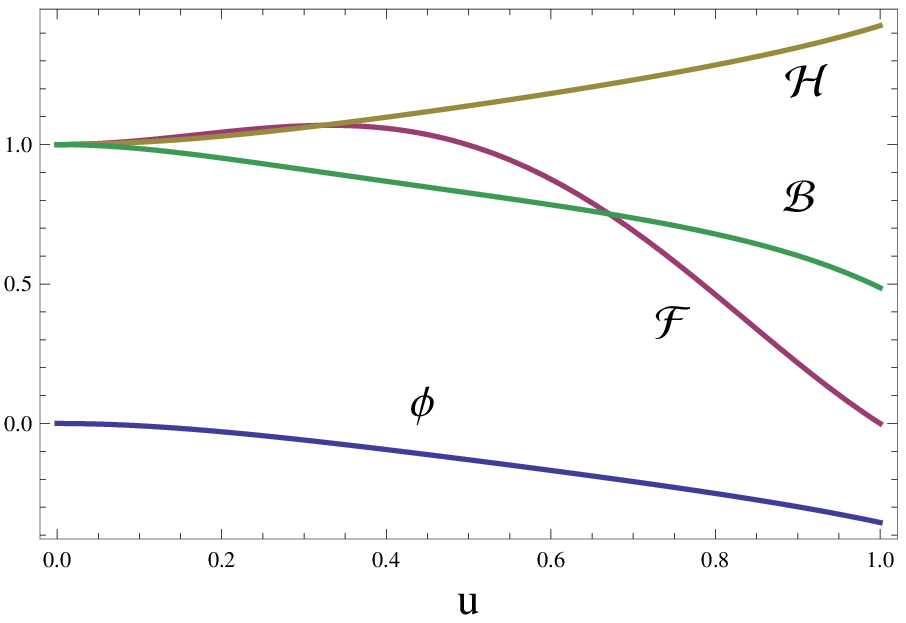}\hspace{16mm}
\includegraphics*[scale=0.50]{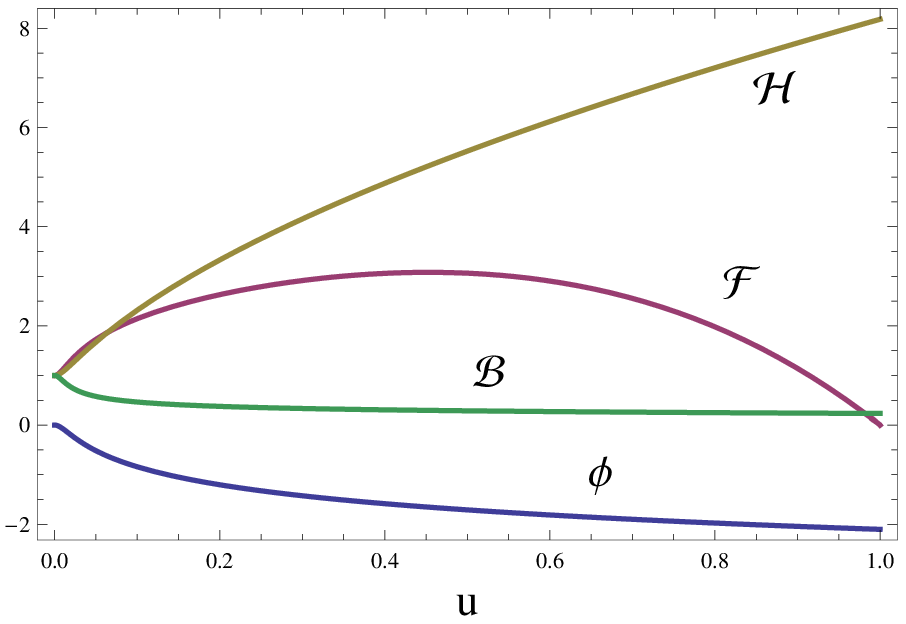}
\end{center}
\caption{(Color online.)The metric functions for $a=1.86$, $Q=6.23$(left) and $a=64.06$, $Q=9.76$ (middle), with $u_H=1$.} \label{metric}
\end{minipage}
\end{figure}

The spacetime is required to be anisotropic but homogenous, so that the functions $\phi$, $\cf$, $\cb$ and $\ch=e^{-\phi}$ should  only depend on the radial coordinate $u$. The electric potential $A_t$ in this metric is expressed  as $A_{t}(u)=-\int^u_{\uh}du Q\sqrt{\cb}e^{\frac{3}{4}\phi}u $ from the Maxwell equations, where $Q$ is an integral constant related to the charge density. Note that the electric potential has a contribution from the anisotropy through the dilatonic field $\phi$.
 The horizon is at $u=\uh$ and the boundary is at $u=0$ respectively. The asymptotic  $AdS$ boundary requires $\cf=\cb=\ch=1$.
We can plot the numerical solutions in Figs.\ref{metric} \cite{ch2} .

In the small anisotropy and charge limits, we can obtain the high temperature solution,
\bea
 &&\cf=1-\bigg(\frac{u}{u_H}\bigg)^4+\bigg[\bigg(\frac{u}{u_H}\bigg)^6-\bigg(\frac{u}{u_H}\bigg)^4\bigg]q^2+a^2 \cf_2(u)+\mathcal{O}(a^4),\nonumber \\[0.7mm]
 &&\cb=1+a^2 \cb_2(u)+\mathcal{O}(a^4),\nonumber \\[0.7mm]
 &&\ch=e^{-\phi(u)}, {~~~\rm with ~~~}  \phi(u)=a^2 \phi_2(u)+\mathcal{O}(a^4),
\label{small_a_exp}
\eea
where $\cf_2(u)=\hat{\cf_0}(u)+\hat{\cf_2}(u) q^2+{O}(q^4),
\cb_2(u)=\hat{\cb_0}(u)+\hat{\cb_2}(u) q^2+{O}(q^4)$ and $
\phi_2(u)=\hat{\phi_0}(u)+\hat{\phi_2}(u)q^2+{O}(q^4)$
with
\bea
\hat\cf_0(u)&=&\frac{1}{24 \uh^2}\left[8 u^2( \uh^2-u^2)-10 u^4\log 2 +(3 \uh^4+7u^4)\log\left(1+\frac{u^2}{\uh^2}\right)\right], \nonumber \\[0.7mm]
\hat\cb_0(u) &=& -\frac{\uh^2}{24}\left[\frac{10 u^2}{\uh^2+u^2} +\log\left(1+\frac{u^2}{\uh^2}\right)\right],                    \nonumber \\[0.7mm]
\hat\phi_0(u)&=& -\frac{\uh^2}{4}\log\left(1+\frac{u^2}{\uh^2}\right)\,.
\eea
and
\bea
\hat\cf_2(u)&=&\frac{1}{24 \uh^4(u^2+\uh^2)} \Big[7u^8+6u^2\uh^6+u^4\uh^4(25\log2-12),\nonumber \\
&&~~~~~~~~~~+u^6\uh^2(25\log2-1)-(u^2+\uh^2)(12u^6+7u^4\uh^2+6\uh^6)\log\left(1+\frac{u^2}{\uh^2}\right) \Big], \nonumber \\[0.7mm]
\hat\cb_2(u) &=& \frac{1}{24}\Big[-\frac{u^2(11u^4+3u^2\uh^2+2\uh^4)}{(u^2+\uh^2)^2}+2\uh^2\log\left(1+\frac{u^2}{\uh^2}\right)\Big]  ,                  \nonumber \\[1.7mm]
\hat\phi_2(u)&=& \frac{1}{4}\Big[-2u^2+\frac{u^4}{u^2+\uh^2}+2\uh^2\log\left(1+\frac{u^2}{\uh^2}\right)\Big].
\eea
where we have used the dimensionless parameter $q=\frac{\uh^3Q}{2\sqrt3}$. Here we would like to justify the usage of the the small anisotropy and charge limits.
The analytic solution for the non-perturbative charge was given in \cite{ch1,ch2}.  Unfortunately,  for the analytic computation of the drag force, it is too difficult to determine
some critical parameters.

 We can easily obtain the temperature as
\bea
T= -\frac{\cf'(\uh) \sqrt{\cb(\uh)}}{4\pi}.
\label{tem}
\eea
Note that the temperature can not be zero unless $a^2\leq0$ which corresponds to the  oblate anisotropy.  In contrast, $a^2>0$ corresponds to the prolate anisotropy. Since  on the dual quantum field theory side, imaginary a looks like a nonunitary deformation and could lead to a negative field coupling. In this sense, the oblate black brane solution with $a^2<0$ could give unphysical results. In the following discussions, we mainly focus on
the prolate case.

\section{Drag force}
On the anisotropic charged brane background, the Nambo-Goto action which governs the dynamics of a probe string in string frame is given by
\bea
S = -T_0 \int d^2 \sigma \, e^{\phi/2}\sqrt{-\det g_{\alpha\beta}},
\eea
where $g_{\alpha\beta} \equiv G_{\mu\nu} \partial_\alpha X^\mu\partial_\beta X^\nu$ is the induced metric on the two-dimensional world sheet. $X^\mu(\sigma^\alpha)$ are the embedding equations of world sheet in spacetime. In what follows, we generalize the calculations in \cite{ma3,gi} to charged plasma case and find the additional contribution to the drag force from the chemical potential. We will work in the static gauge, i.e. $\sigma=u$ and $\tau=t$. Since the plasma is anisotropic, we consider the motions of the string in two different directions $x$ and $z$, which is corresponding to the quark moving at constant velocity $v$. It is convenient to set the
configuration  of string as \cite{ma3}
\bea
x(u,t)&=&x(u)\sin \ta +v t ~ \sin\ta\nonumber \\[0.7mm]
z(u,t)&=&z(u)\cos \ta +v t ~ \cos\ta.
\label{eme}
\eea
Obviously  $\ta$ is the angle between the $z$-axis and the velocity.

The determinant of $g_{\alpha\beta}$ satisfies
\bea
-g e^{\phi}=\frac{-v^2\left(\sin^2\ta+\big(1+\cf \sin^2\ta(x'-z')^2\big)\ch\cos^2\ta\right)+\cb\cf\left(1+\cf(x'^2\sin^2\ta+\ch z'^2\cos^2\ta)\right)}{u^4\cf},
\label{ind}
\eea
so the Lagrangian ${\cal L}=-T_0 e^{\phi/2}\sqrt{-g} $ gives the canonical momenta density to the string as
\bea
\Pi^\sigma_x=\frac{\partial \cal L}{\partial x'}=-T_0e^{-\phi/2}\frac{x'\cb\cf+v^2\ch(z'-x')\cos^2\ta}{u^4\sqrt{-g}}\sin\ta\nonumber \\[0.7mm]
\Pi^\sigma_z=\frac{\partial \cal L}{\partial z'}=-T_0e^{-\phi/2}\frac{z'\cb\cf+v^2\ch(x'-z')\sin^2\ta}{u^4\sqrt{-g}}\cos\ta,
\eea
where the prime denotes the derivative with respect to $u$.
Then the equations of motion following from Nambu-Goto action are
\bea
\partial_u\left(-T_0e^{-\phi/2}\frac{x'\cb\cf+v^2\ch(z'-x')\cos^2\ta}{u^4\sqrt{-g}}\sin\ta \right)=0\nonumber \\[0.7mm]
\partial_u\left(-T_0e^{-\phi/2}\frac{z'\cb\cf+v^2\ch(x'-z')\sin^2\ta}{u^4\sqrt{-g}}\cos\ta \right)=0.
\label{EOM}
\eea
Note that in string configuration (\ref{eme}), the time part of E.O.M is vanish because of time independence. Then after integration of (\ref{EOM}),
we have
\bea
-T_0e^{-\phi/2}\frac{x'\cb\cf+v^2\ch(z'-x')\cos^2\ta}{u^4\sqrt{-g}}\sin\ta=C, \nonumber \\[0.7mm]
-T_0e^{-\phi/2}\frac{z'\cb\cf+v^2\ch(x'-z')\sin^2\ta}{u^4\sqrt{-g}}\cos\ta=D,
\eea
which involve
\bea
&&x'=\frac{e^{\phi/2}u^4\csc\ta \sqrt{-g}\left(v^2(C+D \cot\ta)-C\cb\cf \csc^2\ta\right)}{T_0\cb\cf\left(\cb\cf\csc^2\ta-v^2(1+\ch\cot^2\ta)\right)},\nonumber \\[0.7mm]
&&z'=\frac{e^{\phi/2}u^4\csc\ta \sqrt{-g}\sec\ta\left(v^2\ch(D\csc\ta+C\sec\ta)-D\cb\cf \csc\ta\sec^2\ta\right)}{T_0\cb\cf\ch\left(4\cb\cf\csc^22\ta-v^2(\ch\csc^2\ta+\sec^2\ta)\right)},
\label{cons}
\eea
where the constant $C$ and $D$ are integral constants corresponding to momenta in two directions.
Taking (\ref{cons}) back to (\ref{ind}), we can solve the determinant of induced metric as
\bea
-g e^{\phi}=-\frac{2T_0\cb\ch(\cb\cf-v^2(\cos^2\ta\ch+\sin^2\ta))^2}{I(u)},
\label{cond1}
\eea
with
\bea
I(u)&=&u^4\Big[-2T_0^2\cb^2\cf^2\ch+\cb\cf\left(2D^2u^4+\ch(2C^2u^4+T_0^2v^2+T_0^2v^2\cos2\ta(\ch-1)+T_0^2v^2\ch)\right)\nonumber \\[0.7mm]
&-&2u^4v^2\ch(D\cos\ta+C\sin\ta)^2\Big].
\eea
The positiveness of the  determinant  requires \cite{ma3}
\bea
\cb(u)=\frac{v^2(\ch(u)\cos^2\ta+\sin^2\ta)}{\cf(u)}, \label{cond2}
\eea
at $u=u_c$ where the denominator $I(u)$ vanishes, which implies that the integral constants satisfy
\bea
D=C\ch(u)\cot\ta,
\eea
at $u=u_c$.
So the determinant of induced metric can be simplified as
\bea
-g e^{\phi}=\frac{T_0^2\cb\left(-\cb\cf+v^2(\cos^2\ta\ch+\sin^2\ta)\right)}{-T_0^2u^4\cb\cf+C^2u^8(1+\cot^2\ta\ch)},
\eea
with the choices of
\bea
C=\pm\frac{T_0v \sin\ta}{u_c^2},~~~~~~D=\pm\frac{T_0\ch(u_c)v\cos\ta}{u_c^2}.
\eea
Then the drag force along the $x$-direction and $z$-direction on string are
\bea
F_x=\frac{T_0v \sin\ta}{u_c^2},~~~~F_z=\frac{T_0\ch(u_c)v\cos\ta}{u_c^2}
\label{drag}
\eea
\section{Numerical analysis}

\begin{figure}
 \begin{minipage}{1\hsize}
\begin{center}
\includegraphics*[scale=0.70]{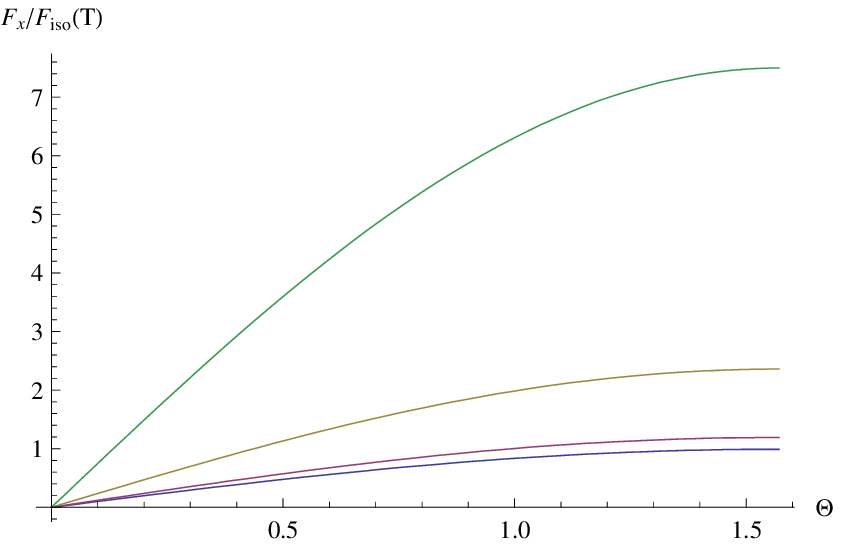}\hspace{4mm}
\includegraphics*[scale=0.70]{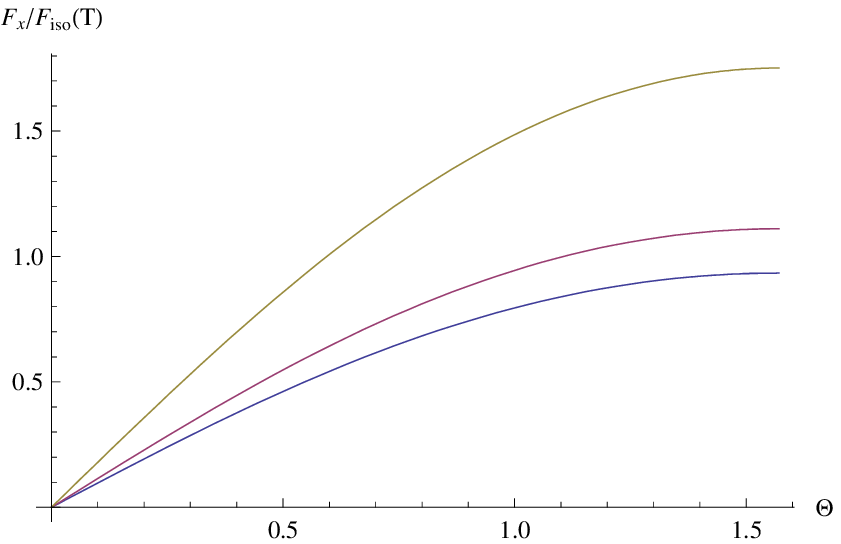}
\includegraphics*[scale=0.70]{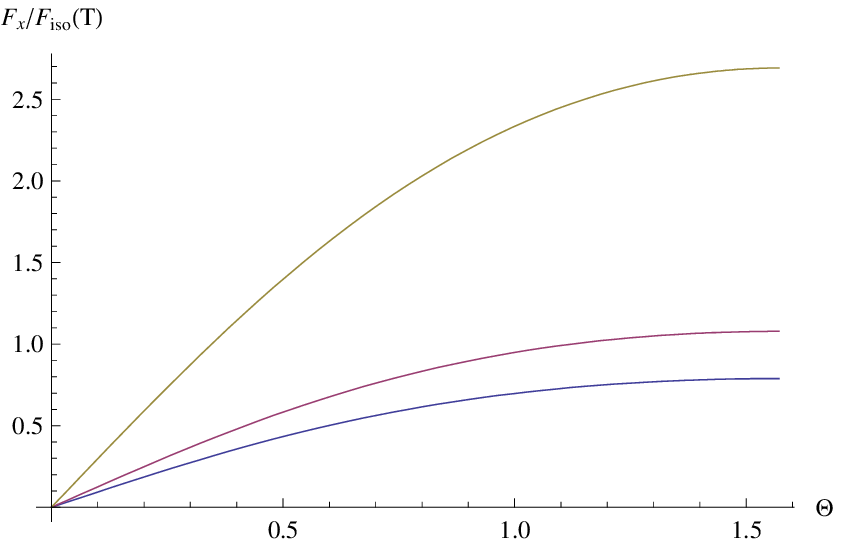}\hspace{4mm}
\includegraphics*[scale=0.70]{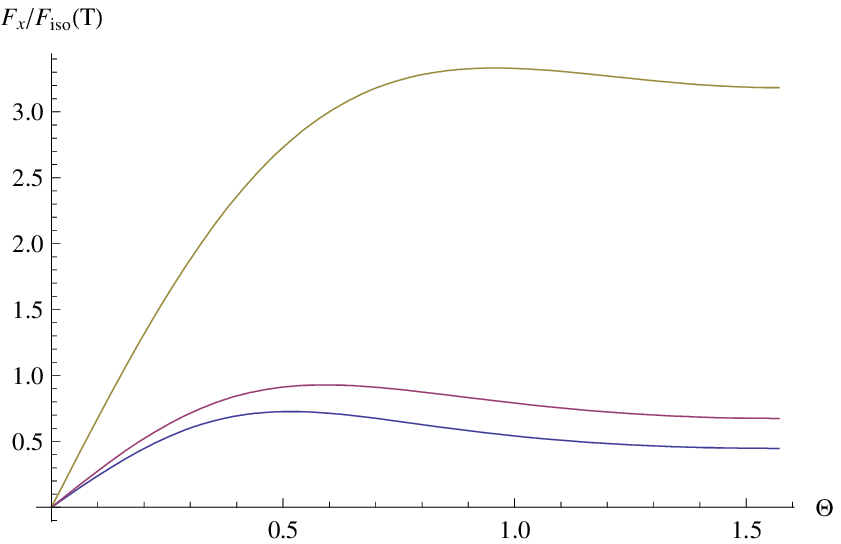}
\end{center}
\caption{Drag force in $x$-direction $F_x$ as a function $\Theta$ at $v=0.5$. The four graphs denote $a/T=1.38$, $a/T=4.41$, $a/T=12.2$, $a/T=86$ respectively,
where the colour lines denote $Q=5$ (Green), $Q=2$(Brown), $Q=1$(Red), $Q=0$(Blue).}  \label{Fx1}
\end{minipage}
\end{figure}

\begin{figure}
 \begin{minipage}{1\hsize}
\begin{center}
\includegraphics*[scale=0.70]{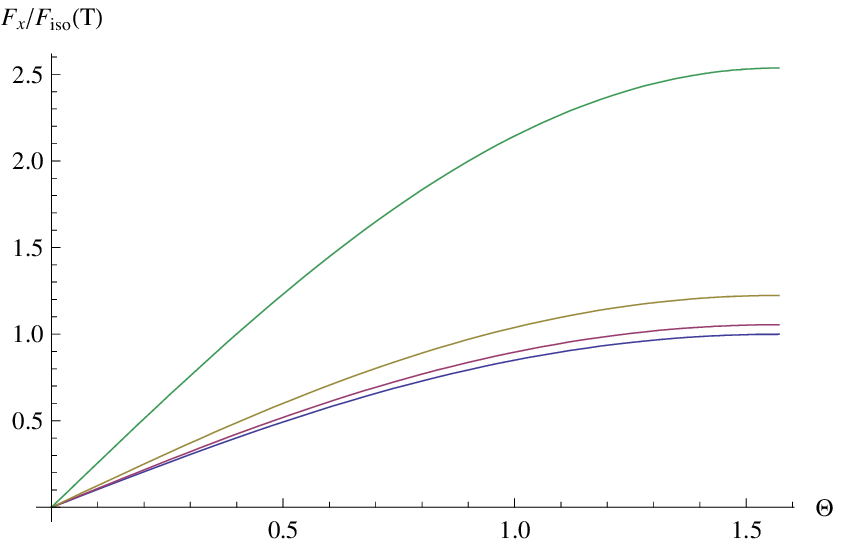}\hspace{4mm}
\includegraphics*[scale=0.70]{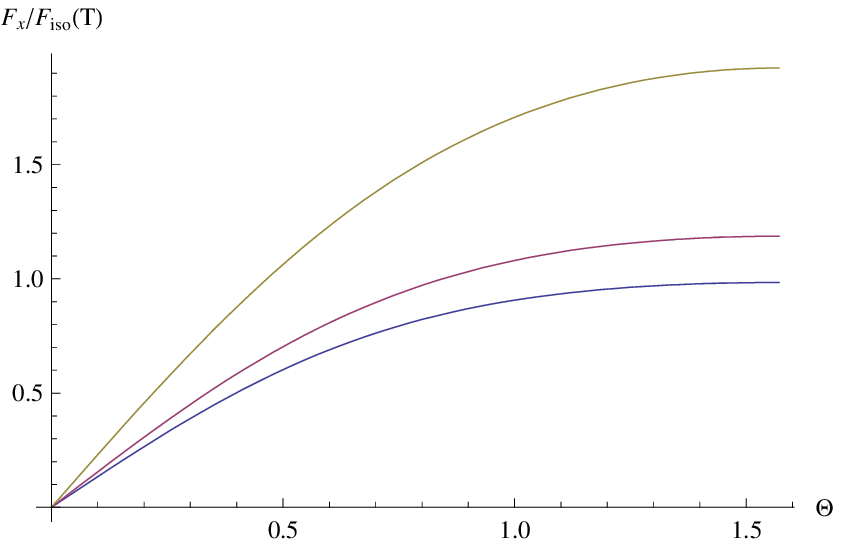}
\includegraphics*[scale=0.70]{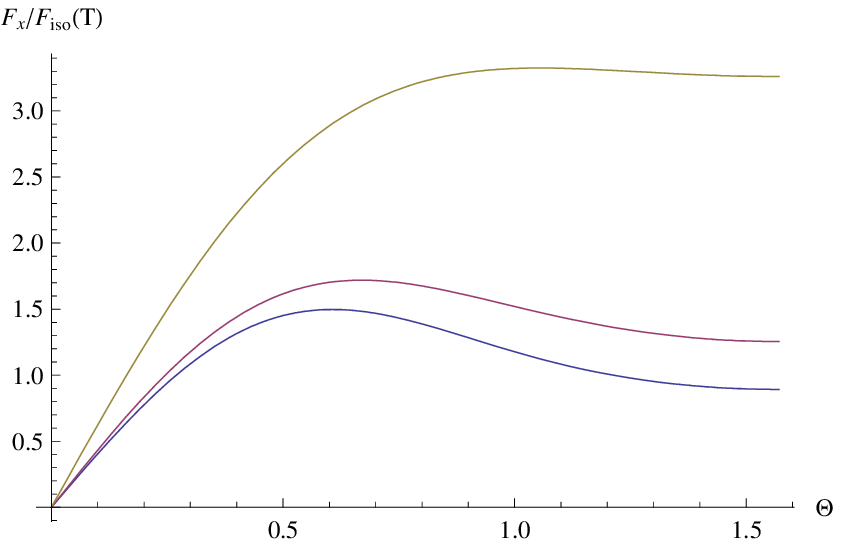}\hspace{4mm}
\includegraphics*[scale=0.70]{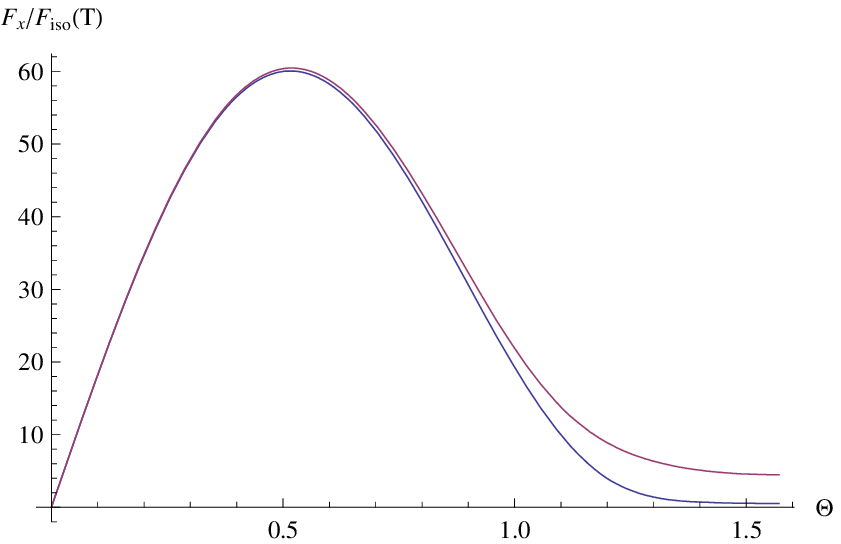}
\end{center}
\caption{(color online)Drag force in $x$-direction $F_x$ as a function $\Theta$ at $v=0.9$. The four graphs denote $a/T=1.38$, $a/T=4.41$, $a/T=12.2$, $a/T=86$ respectively.
where the colour lines denote $Q=5$ (Green), $Q=2$(Brown), $Q=1$(Red), $Q=0$(Blue).}  \label{Fx2}
\end{minipage}
\end{figure}

\begin{figure}
 \begin{minipage}{1\hsize}
\begin{center}
\includegraphics*[scale=0.70]{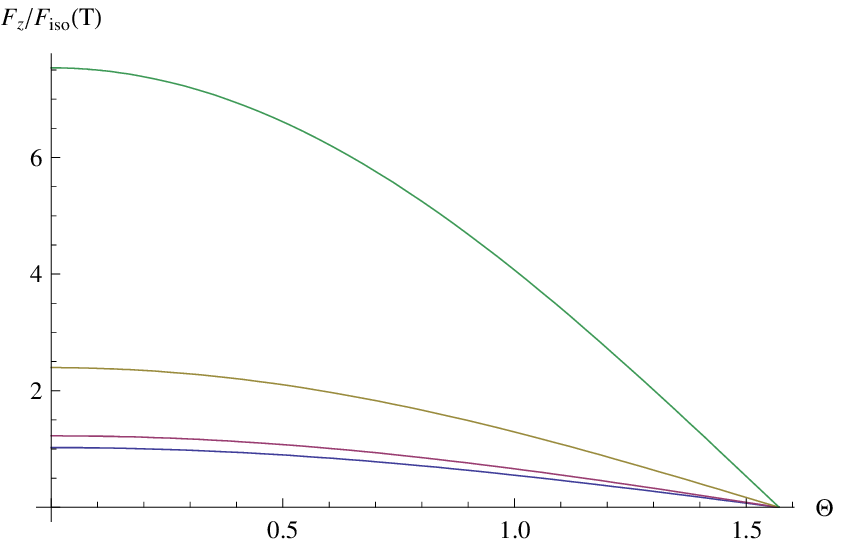}\hspace{4mm}
\includegraphics*[scale=0.70]{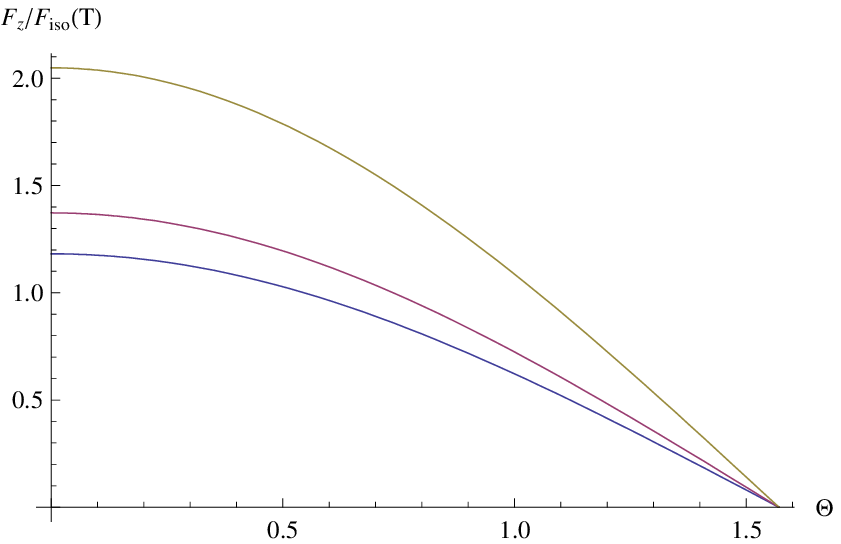}
\includegraphics*[scale=0.70]{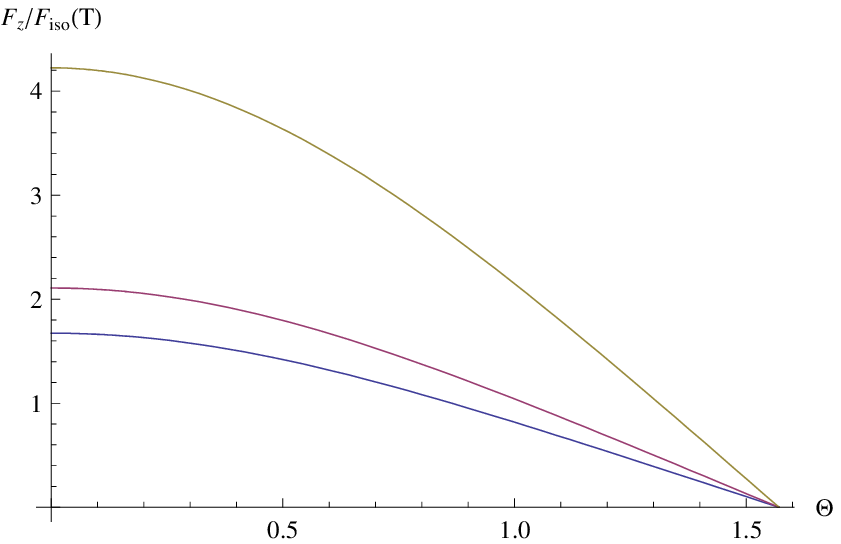}\hspace{4mm}
\includegraphics*[scale=0.70]{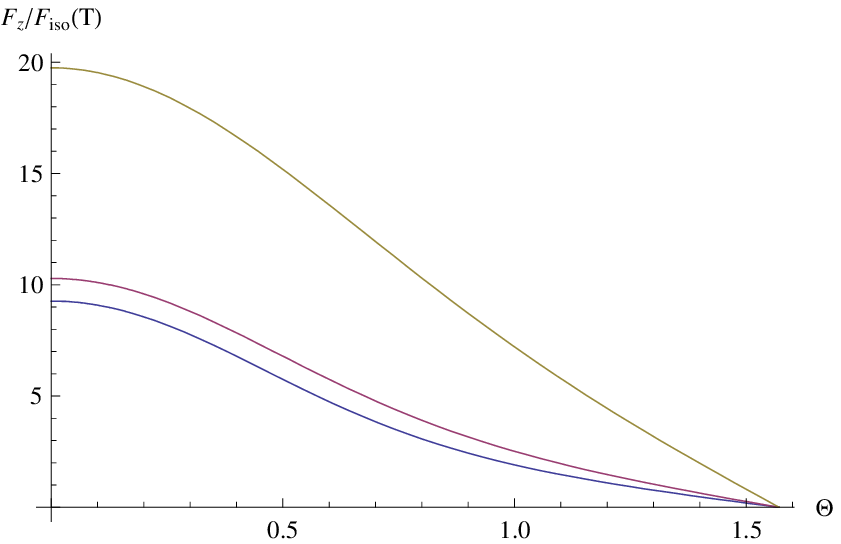}
\end{center}
\caption{Drag force in $z$-direction $F_z$ as a function $\Theta$ at $v=0.5$. The four graphs denote $a/T=1.38$, $a/T=4.41$, $a/T=12.2$, $a/T=86$ respectively,
where the colour lines denote $Q=5$ (Green), $Q=2$(Brown), $Q=1$(Red), $Q=0$(Blue).}  \label{Fz1}
\end{minipage}
\end{figure}

\begin{figure}
 \begin{minipage}{1\hsize}
\begin{center}
\includegraphics*[scale=0.70]{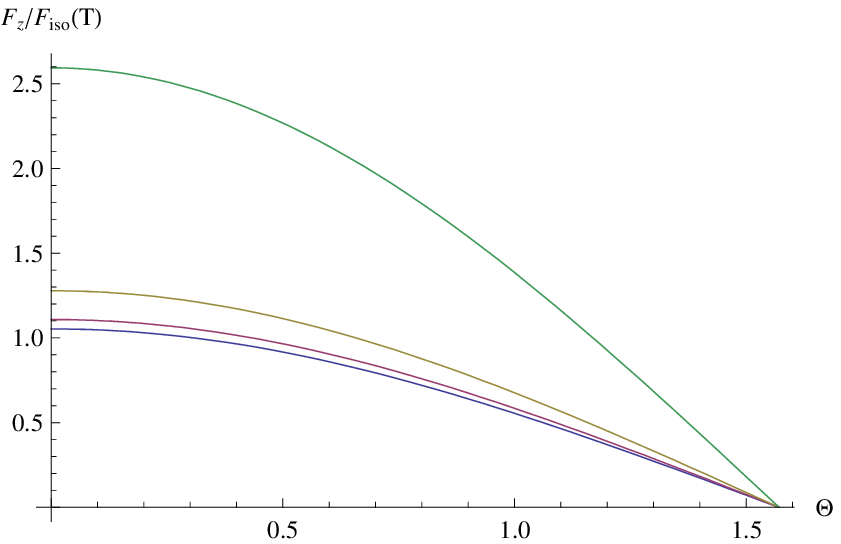}\hspace{4mm}
\includegraphics*[scale=0.70]{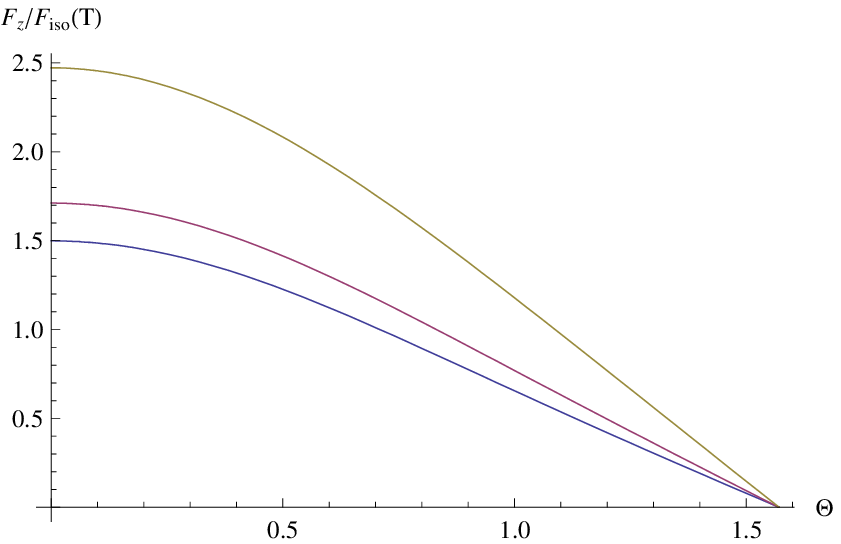}
\includegraphics*[scale=0.70]{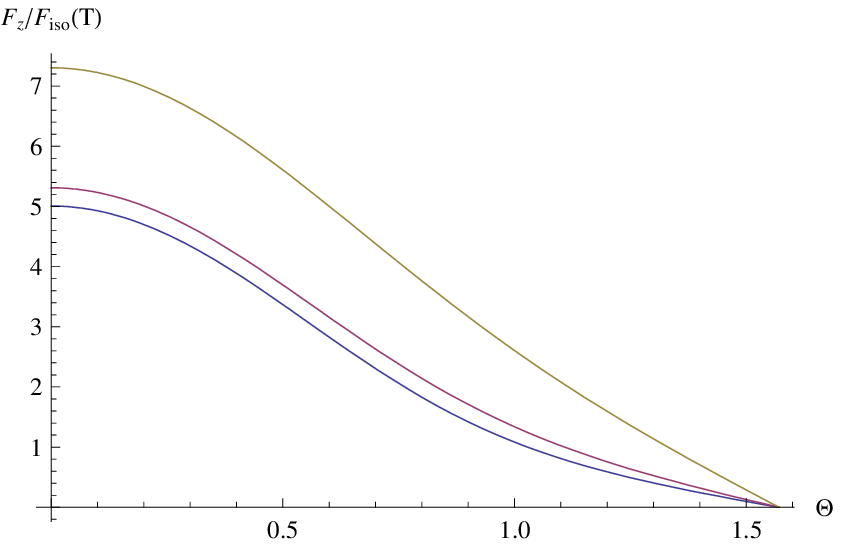}\hspace{4mm}
\includegraphics*[scale=0.70]{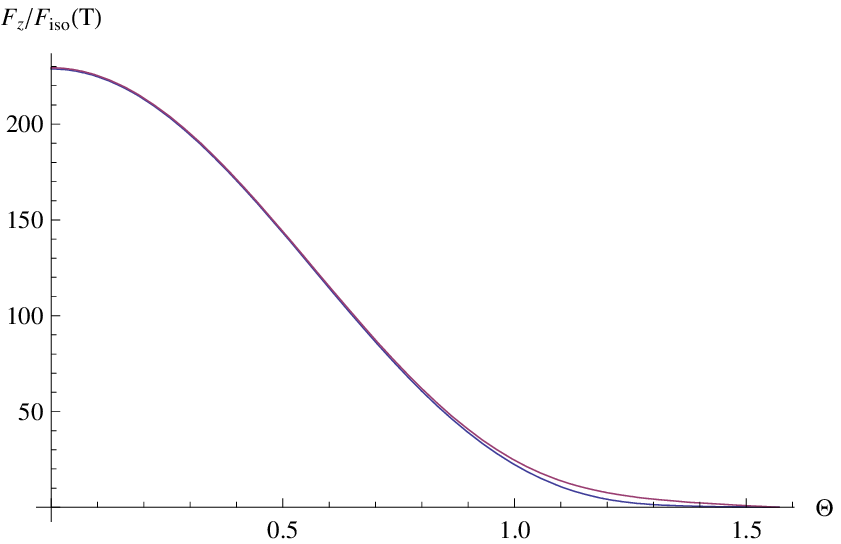}
\end{center}
\caption{Drag force in $z$-direction $F_z$ as a function $\Theta$ at $v=0.9$. The four graphs denote $a/T=1.38$, $a/T=4.41$, $a/T=12.2$, $a/T=86$ respectively,
where the colour lines denote $Q=5$ (Green), $Q=2$(Brown), $Q=1$(Red), $Q=0$(Blue).}  \label{Fz2}
\end{minipage}
\end{figure}

In charged anisotropic plasma, the angle dependence of drag force in $x$-direction  in units of the isotropic drag force are shown in Fig.\ref{Fx1} and Fig.\ref{Fx2}. From the first three plots
of Fig.\ref{Fx1}, we can see that $F_x$ is monotonically increasing functions of  $\Theta$ when $a\ll T$ and $a\sim T$. However, the last plot of Fig.\ref{Fx1}
illustrates that $F_x$ is no longer a monotonic function of $\Theta$ when $a\gg T$. By contrast, Fig.\ref{Fx2} shows that the $F_x$ at $v=0.9$ is no longer a monotonic function of $\Theta$  when $a/T=12.2$. We also find that for charged plasma, with fixed $v$, $F_x$ can be larger than $F_{iso}\equiv F_{iso}(q=0)$ for some interval of $\Theta$ which depend on $Q$ and $a/T$, by contrast, $F_x$ is always smaller than $F_{iso}$ for chargeless anisotropic plasma. Note that  the non-monotonic behavior shown in Fig.\ref{Fx2} is consistent with the result given in \cite{ma3}, where the authors did not show $F_x$. This is because if we write down the general expression for $F$, we can obtain the same ansatz as that given in \cite{ma3} in the zero-chemical potential limit.
Similarly, we can plot the angle dependence of drag force in $z$-direction in Fig.\ref{Fz1} and Fig.\ref{Fz2}, while $F_z$ is always the monotonically decreasing function of $\Theta$.

\begin{figure}
 \begin{minipage}{1\hsize}
\begin{center}
\includegraphics*[scale=0.70]{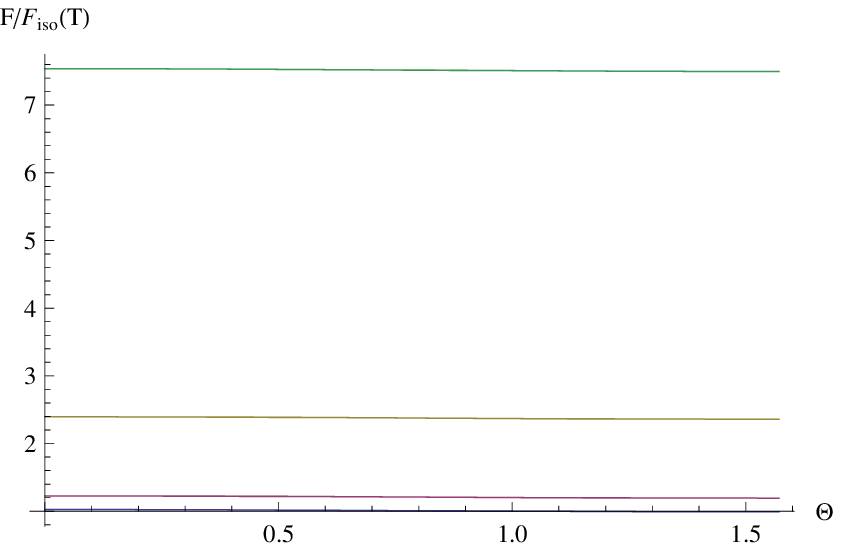}\hspace{4mm}
\includegraphics*[scale=0.70]{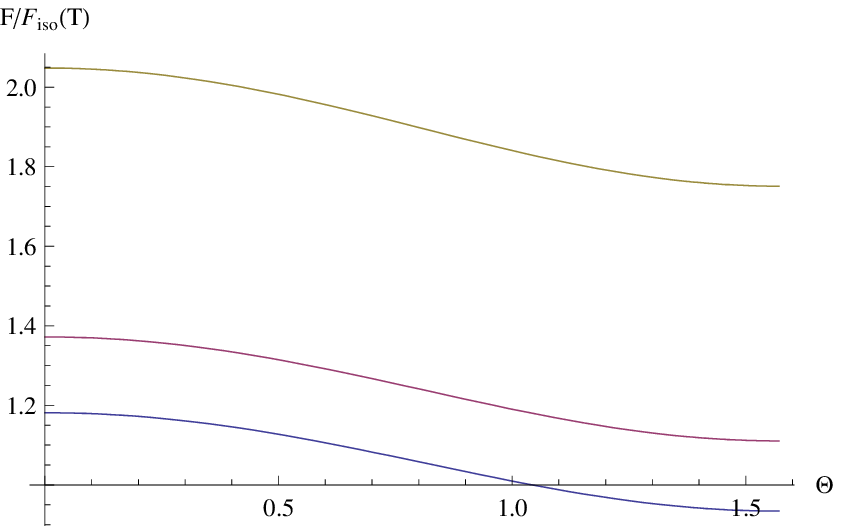}
\includegraphics*[scale=0.70]{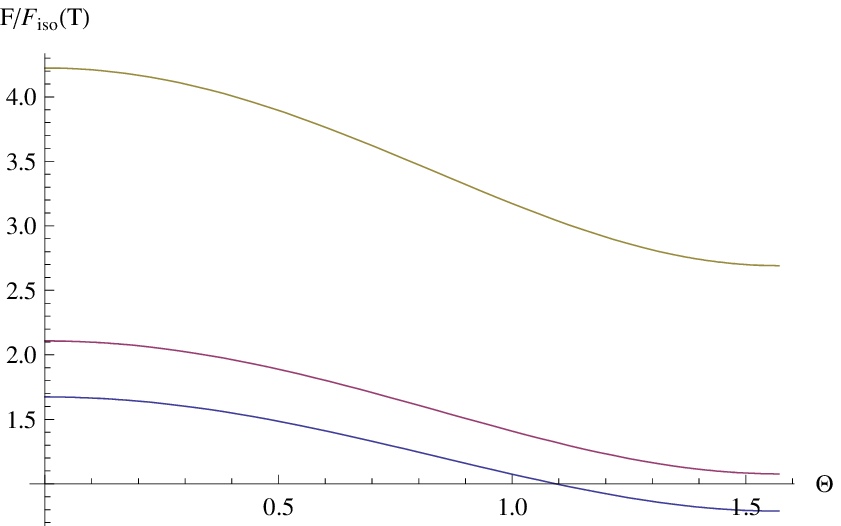}\hspace{4mm}
\includegraphics*[scale=0.70]{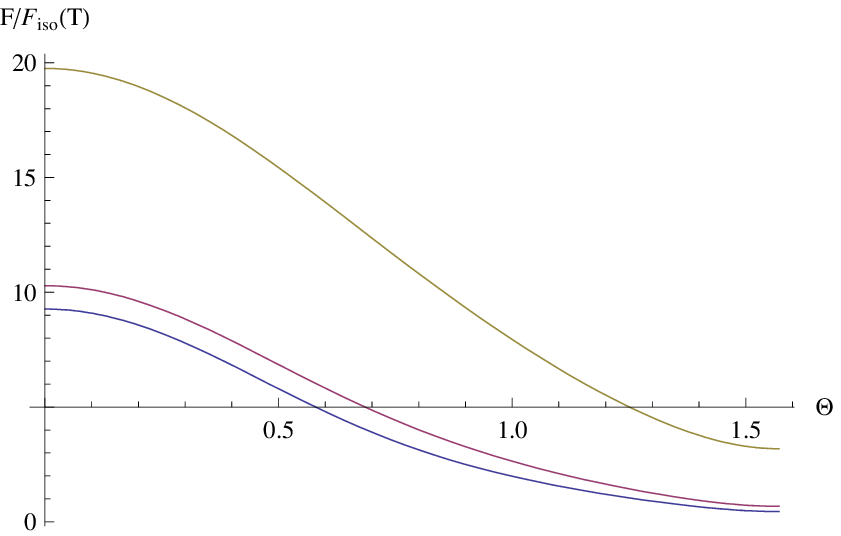}
\end{center}
\caption{Drag force $F$ as a function $\Theta$ at $v=0.5$. The four graphs denote $a/T=1.38$, $a/T=4.41$, $a/T=12.2$, $a/T=86$ respectively,
where the colour lines denote $Q=5$ (Green), $Q=2$(Brown), $Q=1$(Red), $Q=0$(Blue).}  \label{F1}
\end{minipage}
\end{figure}

\begin{figure}
 \begin{minipage}{1\hsize}
\begin{center}
\includegraphics*[scale=0.70]{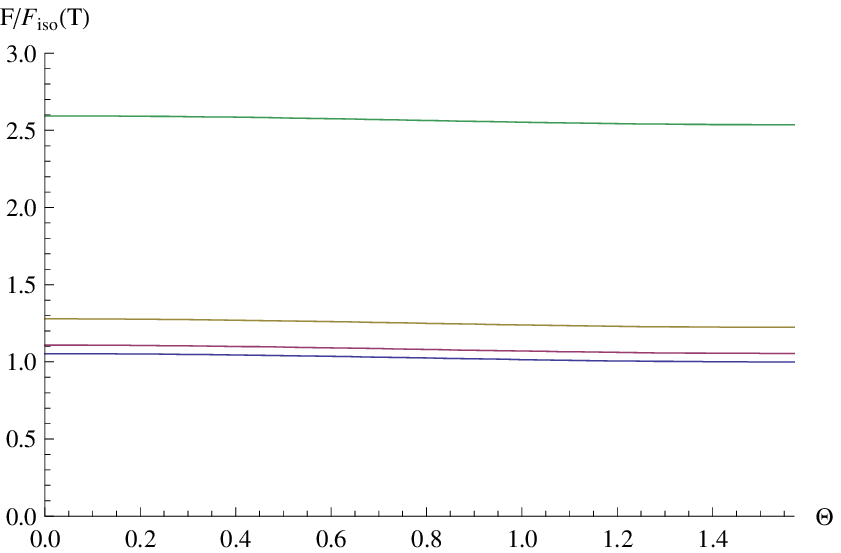}\hspace{4mm}
\includegraphics*[scale=0.70]{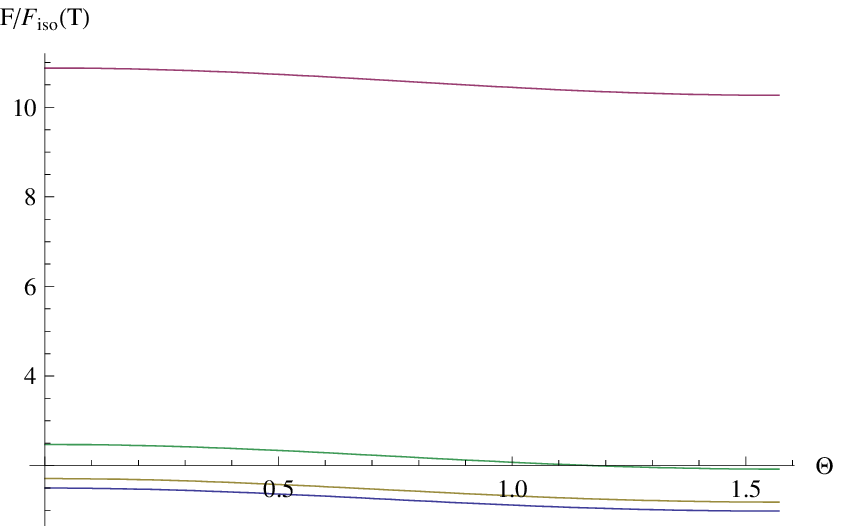}
\includegraphics*[scale=0.70]{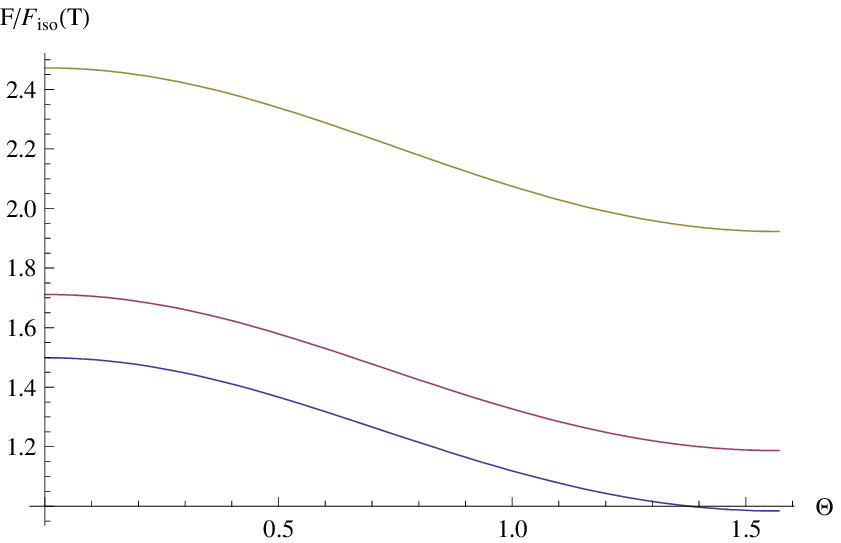}\hspace{4mm}
\includegraphics*[scale=0.70]{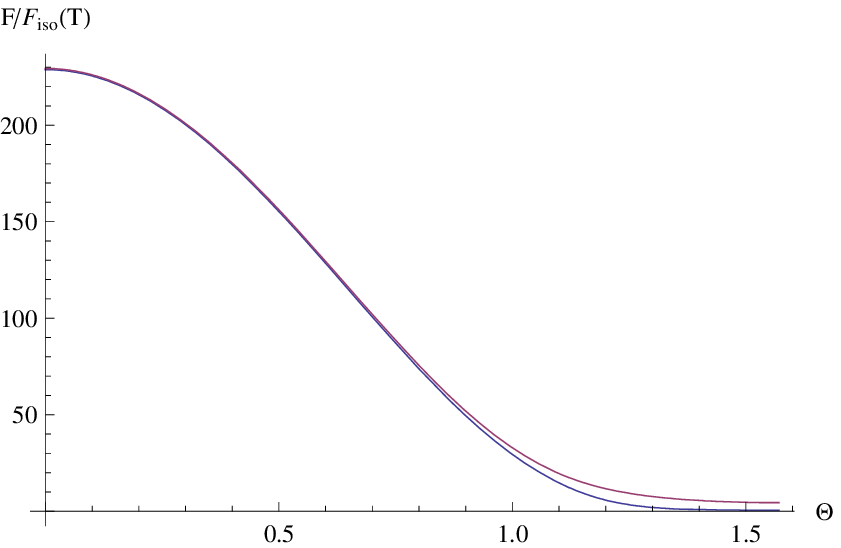}
\end{center}
\caption{(color online)Drag force $F$ as a function $\Theta$ at $v=0.9$. The four graphs denote $a/T=1.38$, $a/T=4.41$, $a/T=12.2$, $a/T=86$ respectively,
where the colour lines denote $Q=5$ (Green), $Q=2$(Brown), $Q=1$(Red), $Q=0$(Blue).}  \label{F2}
\end{minipage}
\end{figure}

The Fig.\ref{F1} and Fig.\ref{F2} illustrate the $\Theta$ dependence of the drag force $F$ in units of the isotropic drag force at fixed $v$ and $a/T$. It is easy to see that $F$ is  monotonically decreasing functions of $\Theta$, and when $a/T$ is greater, the falloff of $F$ is faster.

\begin{figure}
\begin{minipage}{1\hsize}
\begin{center}
\includegraphics*[scale=0.70] {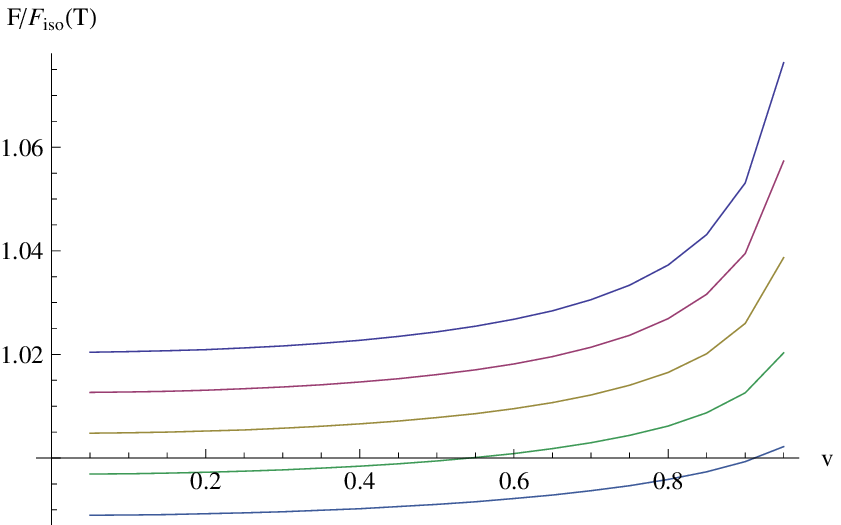}\hspace{8mm}\hspace{4mm}
\includegraphics*[scale=0.70] {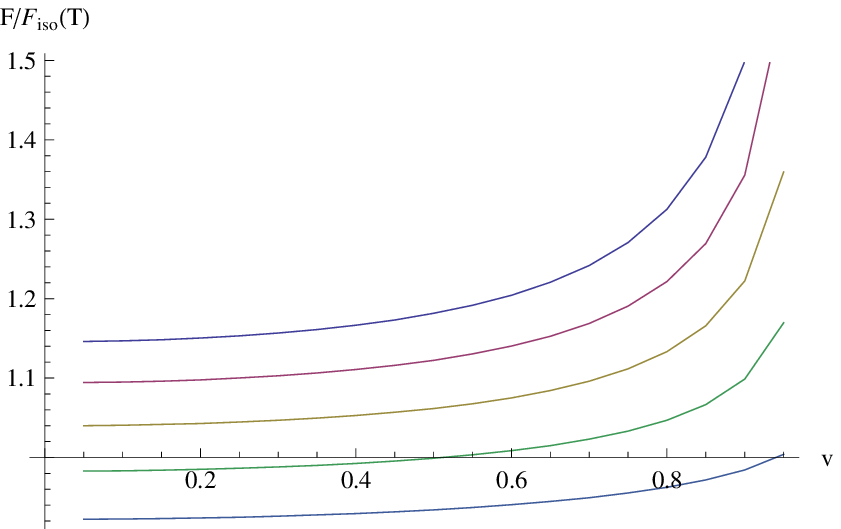}\vspace{8mm}
\includegraphics*[scale=0.70] {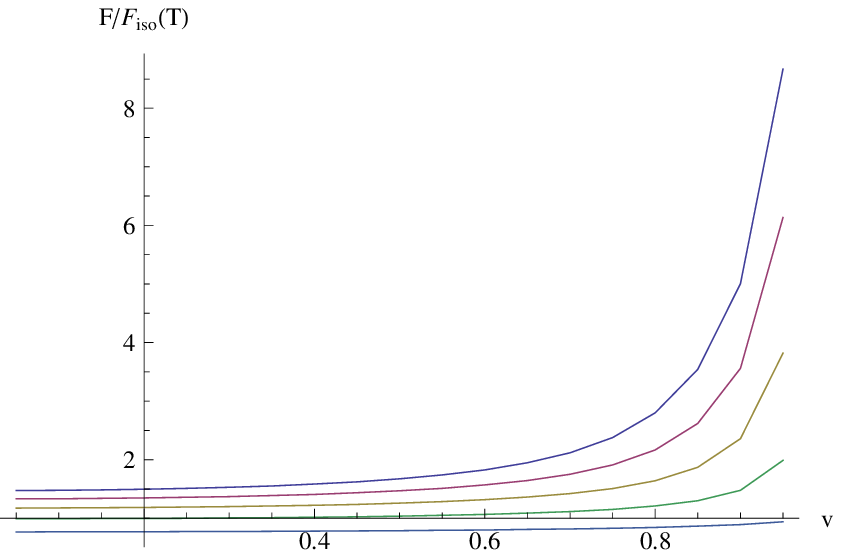}\hspace{8mm}\hspace{4mm}
\includegraphics*[scale=0.70] {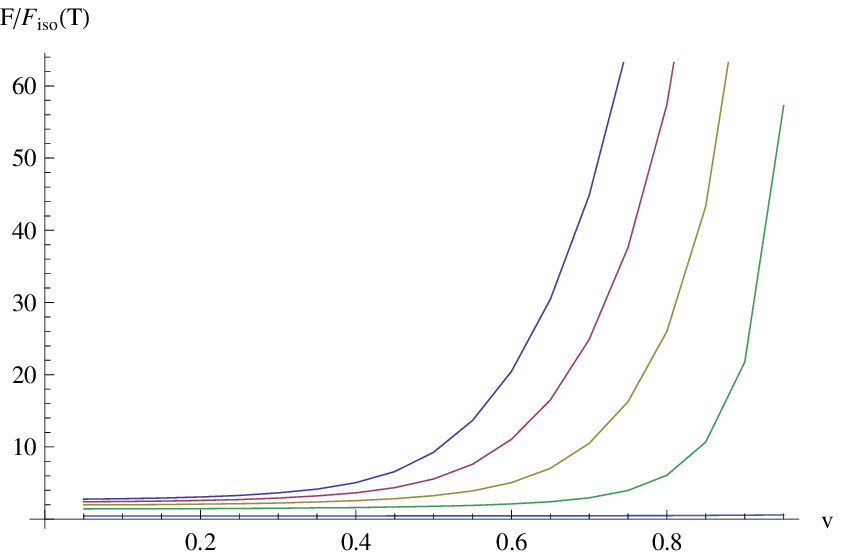}\hspace{8mm}
\end{center}
\caption{The drag force $F$ as a function of velocity for a quark moving through the plasma with $Q=0$ (i.e. chargeless). The four graphs correspond to $a/T=1.38,
 a/T=4.42, a/T=12.2, a/T=86$ respectively, in which five lines denotes (from the top to down ) $\Theta=0$, $\Theta=\pi/6$, $\Theta=\pi/4$, $\Theta=\pi/3$, $\Theta=\pi/2$}\label{F-v0}
\end{minipage}
\end{figure}

\begin{figure}
 \begin{minipage}{1\hsize}
\begin{center}
\includegraphics*[scale=0.70] {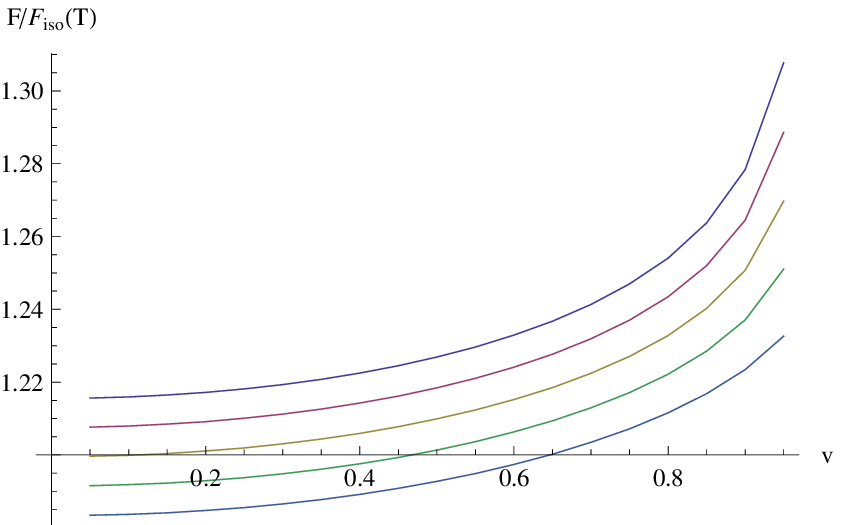}\hspace{8mm}\hspace{4mm}
\includegraphics*[scale=0.70] {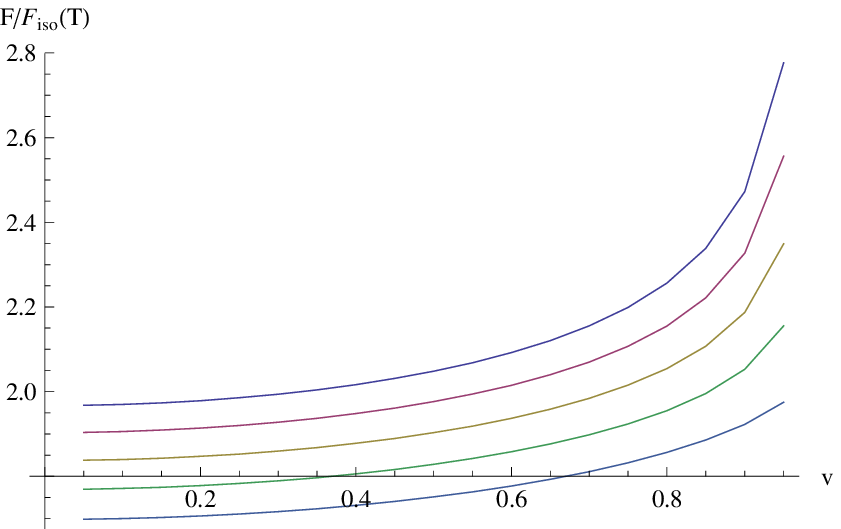}\vspace{8mm}
\includegraphics*[scale=0.70] {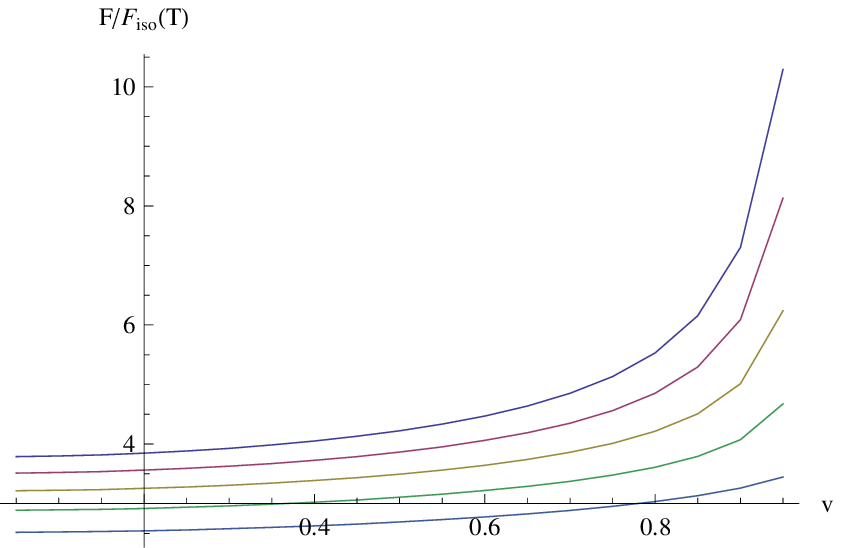}\hspace{8mm}\hspace{4mm}
\includegraphics*[scale=0.70] {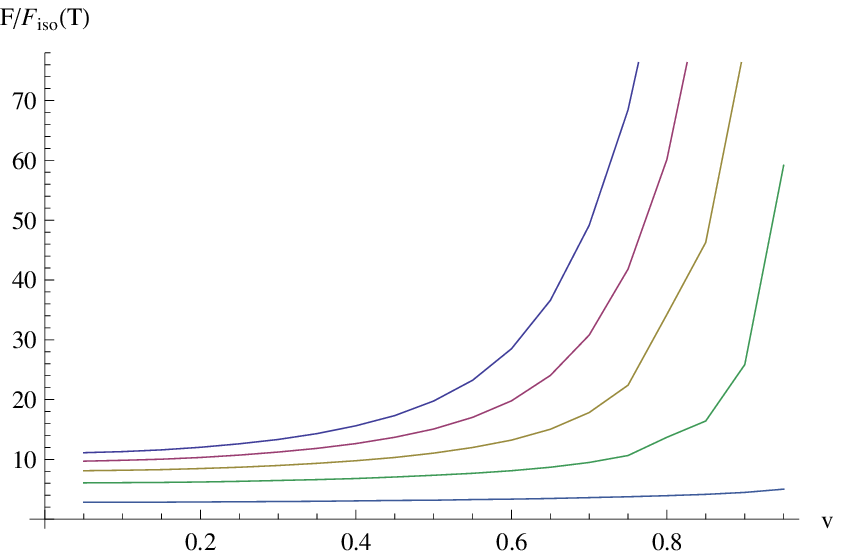}\hspace{8mm}
\end{center}
\caption{The drag force $F$ as a function of velocity for a quark moving through the plasma with $Q=2$. The four graphs correspond to $a/T=1.38,
a/T=4.42,a/T=12.2,a/T=86$ respectively, in which five lines denotes (from the top to down ) $\Theta=0$, $\Theta=\pi/6$, $\Theta=\pi/4$, $\Theta=\pi/3$,$\Theta=\pi/2$}
\label{F-v2}
\end{minipage}
\end{figure}

\begin{figure}
 \begin{minipage}{1\hsize}
\begin{center}
\includegraphics*[scale=0.70] {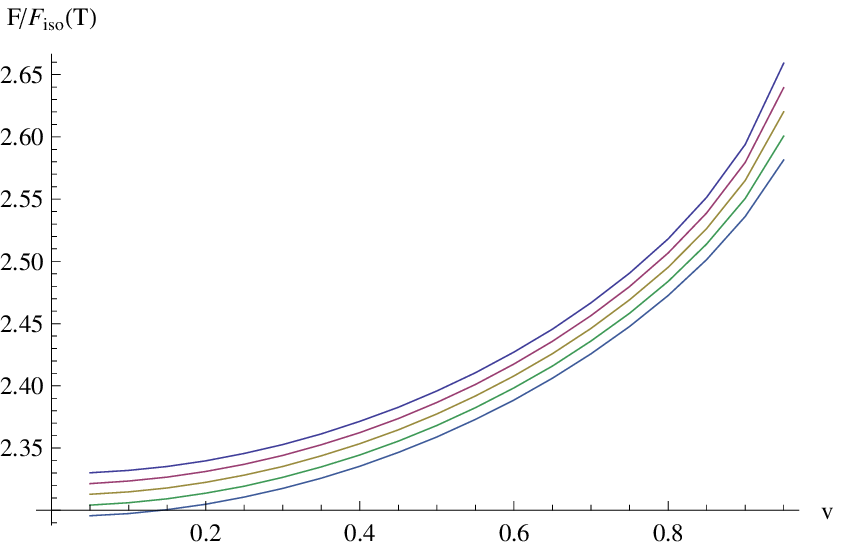}\hspace{8mm}\hspace{4mm}
\includegraphics*[scale=0.70] {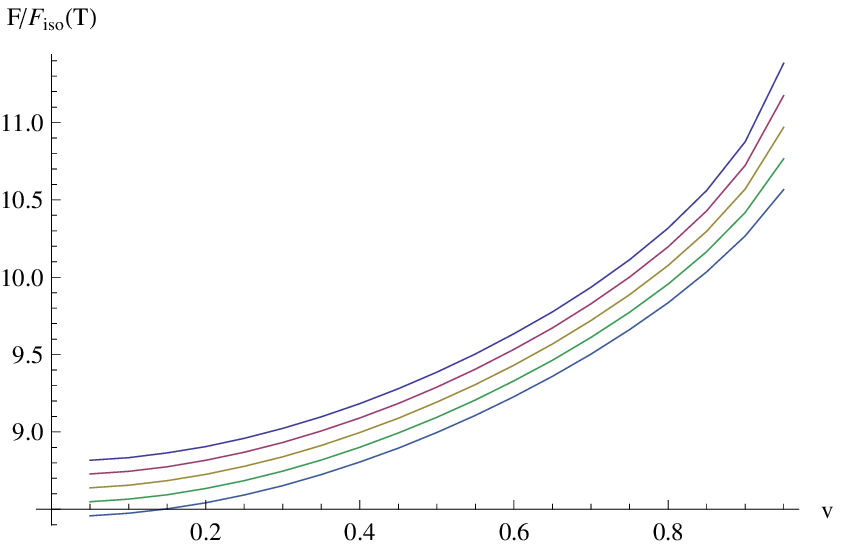}\vspace{8mm}
\includegraphics*[scale=0.70] {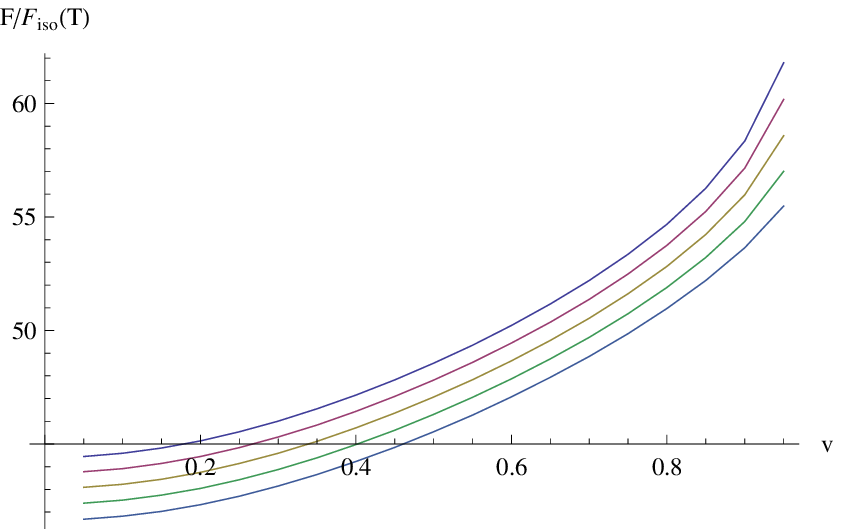}\hspace{8mm}\hspace{4mm}
\includegraphics*[scale=0.70] {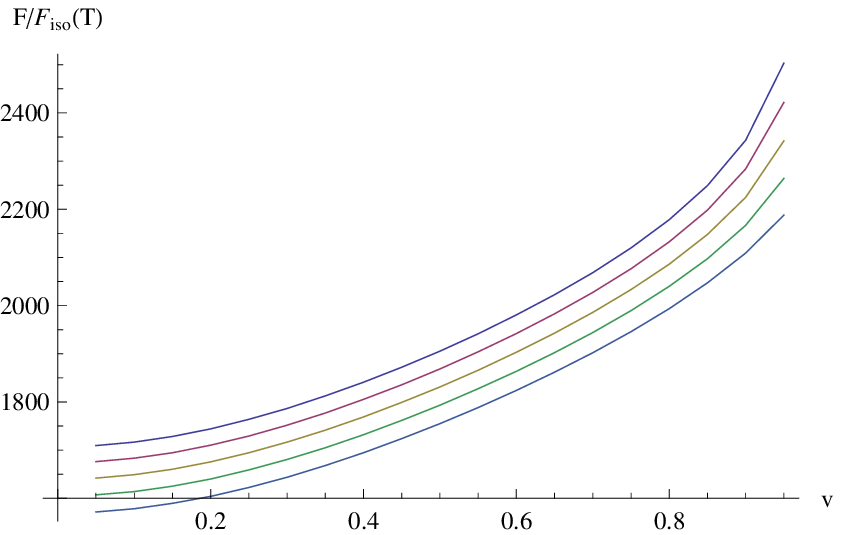}\hspace{8mm}
\end{center}
\caption{The drag force $F$ as a function of velocity for a quark moving through the plasma with $Q=5$. The four graphs correspond to $a/T=1.38,
a/T=4.42,a/T=12.2,a/T=86$ respectively, in which five lines denotes (from the top to down ) $\Theta=0$, $\Theta=\pi/6$, $\Theta=\pi/4$, $\Theta=\pi/3$,$\Theta=\pi/2$}\label{F-v5}
\end{minipage}
\end{figure}

\begin{figure}
 \begin{minipage}{1\hsize}
\begin{center}
\includegraphics*[scale=0.70] {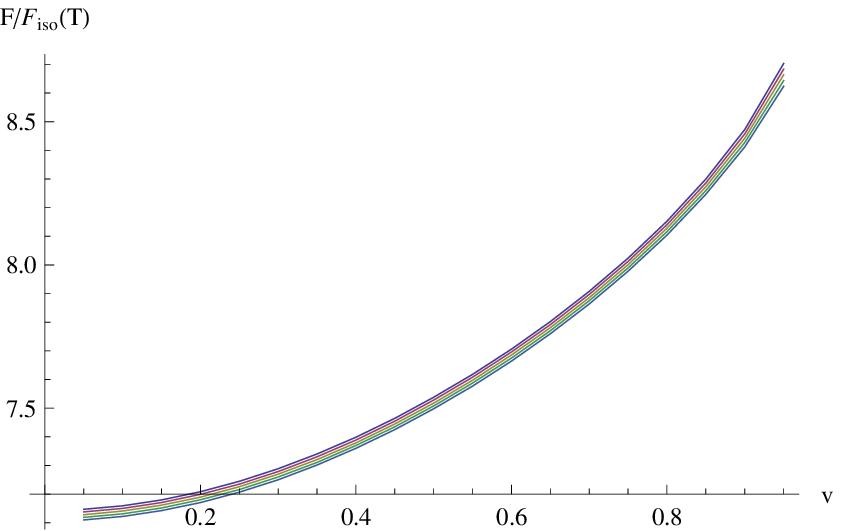}\hspace{8mm}\hspace{4mm}
\includegraphics*[scale=0.70] {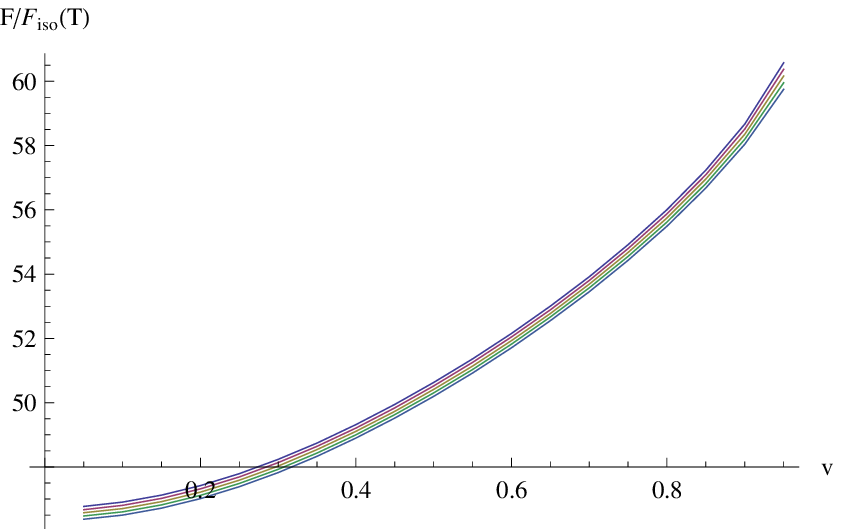}\vspace{8mm}
\includegraphics*[scale=0.70] {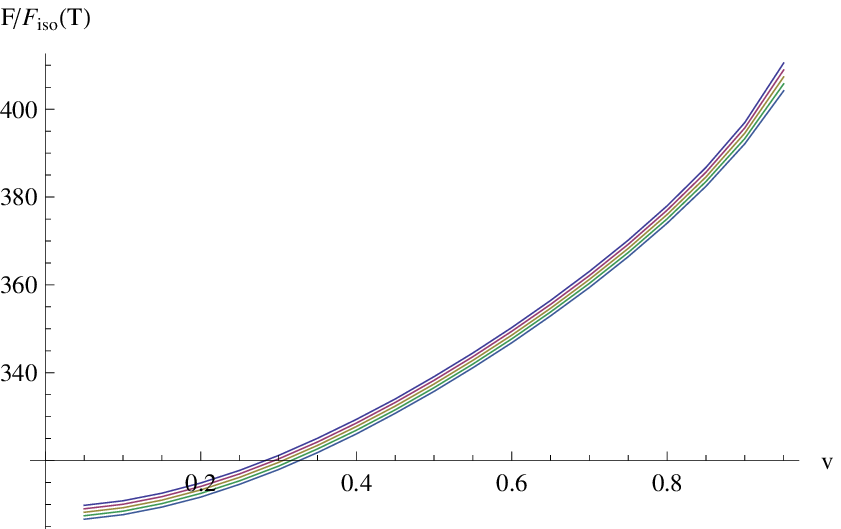}\hspace{8mm}\hspace{4mm}
\includegraphics*[scale=0.70] {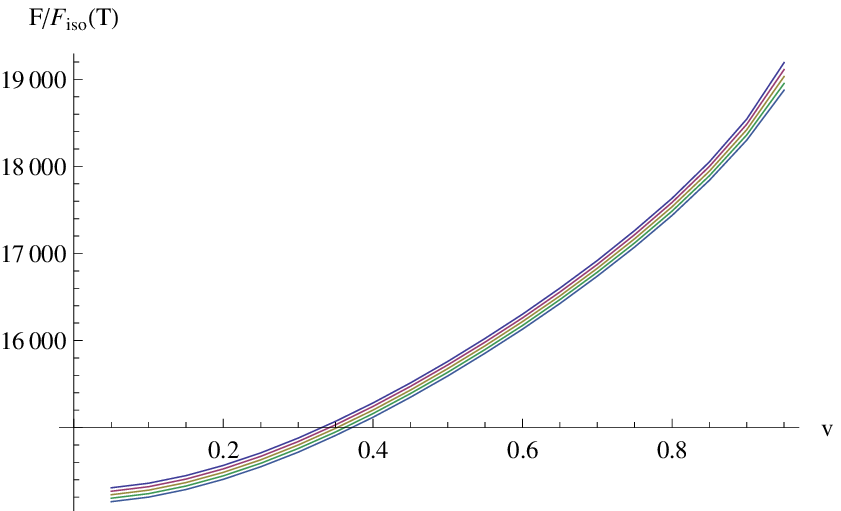}\hspace{8mm}
\end{center}
\caption{The drag force $F$ as a function of velocity for a quark moving through the plasma with $Q=10$. The four graphs correspond to $a/T=1.38,
a/T=4.42,a/T=12.2,a/T=86$ respectively, in which five lines denote (from the top to down ) $\Theta=0$, $\Theta=\pi/6$, $\Theta=\pi/4$, $\Theta=\pi/3$, $\Theta=\pi/2$}\label{F-v10}
\end{minipage}
\end{figure}

In Fig.\ref{F-v0}-Fig.\ref{F-v10}, we can see that the drag force $F$  in units of the isotropic drag force diverges
in the ultra-relativistic limit $v\rightarrow1$ for any $\Theta\neq\pi/2$. We can see that $F$ in charged plasma  diverges faster than in chargeless plasma.
So the anisotropic drag force is arbitrarily larger than the isotropic case in the ultra-relativistic limit. More interestingly, in the large Q limit, the behavior of drag force $F$ coincides for different angle $\Theta$.

\begin{figure}
 \begin{minipage}{1\hsize}
\begin{center}
\includegraphics*[scale=0.70] {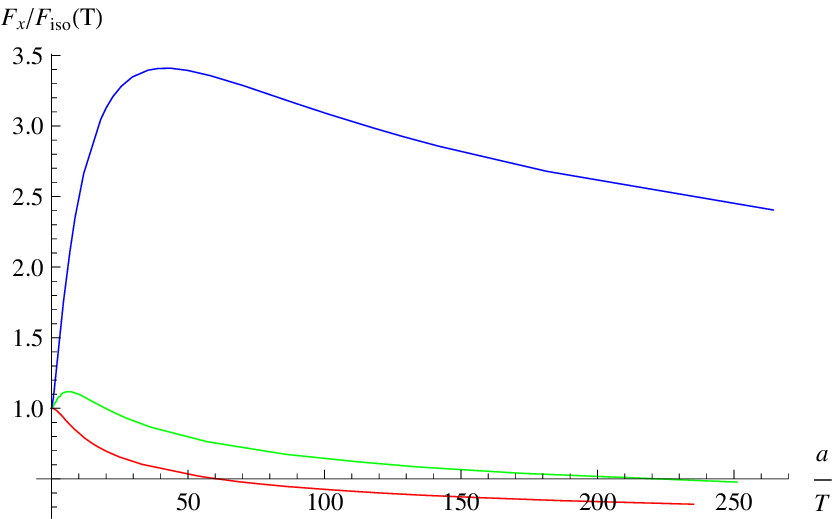}\hspace{4mm}
\includegraphics*[scale=0.70] {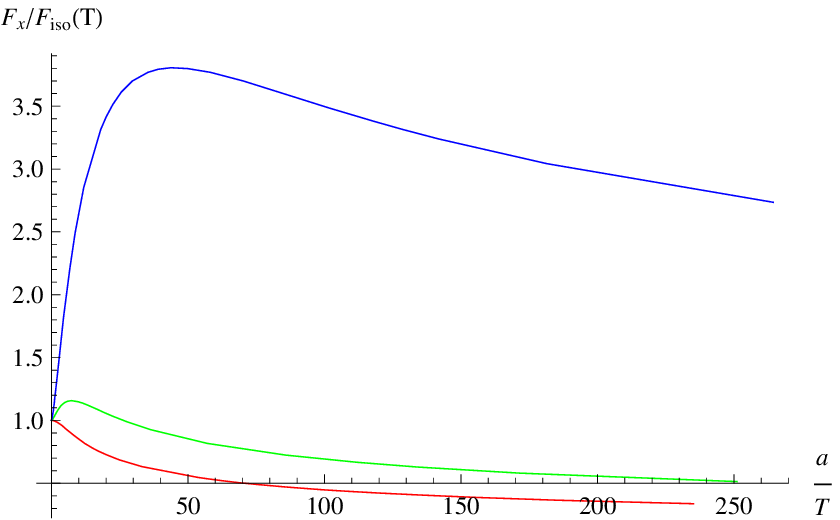}
\end{center}
\caption{Drag force in $x$ direction as a function $a/T$ at $v=0.5$ (left) and $v=0.7$ (right), where the Red, Green, Blue lines represent $Q=0$, $Q=1$, $Q=2$ respectively.} \label{Fx-T}
\end{minipage}
\end{figure}

\begin{figure}
 \begin{minipage}{1\hsize}
\begin{center}
\includegraphics*[scale=0.70] {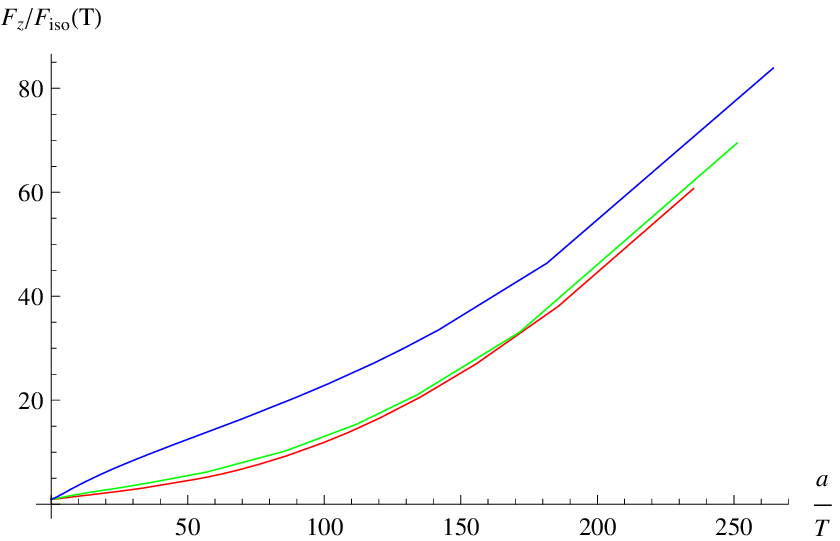}\hspace{4mm}
\includegraphics*[scale=0.70] {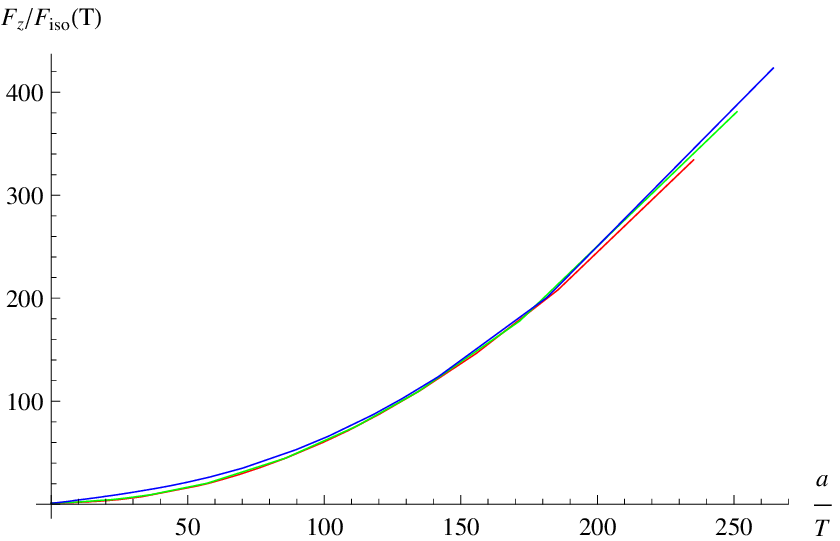}
\end{center}
\caption{Drag force in $z$ direction as a function $a/T$ at $v=0.5$ (left) and $v=0.7$ (right), where the Red, Green, Blue lines represent $Q=0$, $Q=1$, $Q=2$ respectively.} \label{Fz-T}
\end{minipage}
\end{figure}

To exhibit the temperature-dependence of drag force more explicitly, we plot the drag force as function of $a/T$ with different $Q$ in Fig.\ref{Fx-T} and Fig.\ref{Fz-T}. We can see that, unlike anisotropic neutral plasma, in which the drag force in the transverse direction $F_x$ ($\Theta=\frac{\pi}{2}$) is  monotonically decreasing function of $a/T$, $F_x$ in anisotropic charged plasma is no longer a the monotonically  function: In the region with $a/T < 1$, $F_x$ increase as $a/T$ increases, but in the region $a/T > 1$,  $F_x$ decreases as $a/ T$ increases. This means that for non-zero charge $Q$, the force changes non-monotonically. However, as shown in the plot of the $F_z$ ($\Theta=0$) in Fig.\ref{Fz-T}, the drag force along the longitudinal direction is always the monotonically increasing function for both neutral plasma and charged plasma.


\section{Summary}
By using AdS/CFT correspondence we have performed the calculations of drag force exerted on a massive quark moving through a charged, anisotropic ${\cal{N}}=4$ SU(N) Super
Yang-Mills plasma. We used the anisotropic charged black brane solution, which is dual to anisotropic QGP with a chemical potential.
For a complete study of the drag force in the anisotropic background, we carried out analytic calculation first and obtain some general expressions for the drag force. Different from the isotropic case, where the drag force in charged plasma  is always larger than in neutral plasma at the same temperature, for our  anisotropic case, we will find that it will be dependent on the explicit value of $q$ and $a$, which can been seen in the numerical analysis. For arbitrary anisotropy and charge, we presented the numerical results for any prolate  anisotropy and arbitrary direction of the quark velocity with respect to the direction of anisotropy. We find the effect of chemical potential or charge density will enhance the drag force for the prolate solution.

\section*{Acknowledgments}
XHG was partly supported by NSFC,
China (No.11375110). SYW was supported by Ministry of Science and Technology and the National Center of Theoretical Science in Taiwan (grant no. MOST-101-2112-M-009-005)

\appendix

\section{High temperature analysis}
In this appendix, we calculate the drag forces in the limit of high temperature.
 We consider the high-temperature solution (\ref{small_a_exp}). First, plugging the (\ref{small_a_exp}) into (\ref{cond2}), it is easy to obtain $u_c$ to order of $a^2$:
\bea
u_c=u_0+u_1a^2,
\eea
with
\bea
&&u_0=\left(\uh^2\sqrt{1-v^2}-\frac{\uh^2(-1+v^2+\sqrt{1-v^2})}{2}q^2\right)^{1/2},\nonumber\\
&&u_1=\frac{\uh^3\left(1-\sqrt{1-v^2}-5\log2+5\log(1+\sqrt{1-v^2})-v^2(1-5\log2+(4+3\cos^2\ta)\log(1+\sqrt{1-v^2}))\right)}{48(1-v^2)^{3/4}}\nonumber\\
&&+\frac{\uh^3}{384(1-v^2)^{5/4}(1+\sqrt{1-v^2})}\Big(10(1-v^2)(8(1+\sqrt{1-v^2})\log2-v^2(8\sqrt{1-v^2}+(8-7\sqrt{1-v^2})\log2))\nonumber\\
&&+(-80(1+\sqrt{1-v^2})+v^4(-86+13\sqrt{1-v^2})+2v^2(83+38\sqrt{1-v^2}))\log(1+\sqrt{1-v^2})\nonumber\\
&&+9v^2\cos2\ta(2(v^2-1)(1+\sqrt{1-v^2})+(v^2(-2+\sqrt{1-v^2})+2(1+\sqrt{1-v^2}))\log(1+\sqrt{1-v^2}))\Big)q^2.\nonumber\\
\eea
So, by use of (\ref{tem}) and  (\ref{drag}), we are able to derive the drag force
\bea
F_x&=&\frac{\pi\sqrt{\lambda}T^2v}{2}\left(\sqrt{\frac{1}{1-v^2}}-\frac{q^2}{2}\big(1-\frac{3}{\sqrt{1-v^2}}\big)\right)+\frac{v\sqrt{\lambda}a^2}{48\pi}\Big(\frac{1-v^2+\sqrt{1-v^2}+(4v^2-5)\log(1+\sqrt{1-v^2})}{(1-v^2)^{3/2}}\nonumber\\
&&+\frac{q^2}{2(1-v^2)^2(1+\sqrt{1-v^2})}\big(-(1+\sqrt{1-v^2})(13+30\log2)+v^2(23+29\sqrt{1-v^2}\nonumber\\
&&+15(4+\sqrt{1-v^2})\log2)+v^4(15\sqrt{1-v^2}\log2-2(5+8\sqrt{1-v^2}+15\log2))\nonumber\\
&&+(20(1+\sqrt{1-v^2})+v^4(17+5\sqrt{1-v^2})-v^2(37+22\sqrt{1-v^2}))\log(1+\sqrt{1-v^2})\big)\Big),\nonumber\\
F_z&=&\frac{\pi\sqrt{\lambda}T^2v}{2}\left(\sqrt{\frac{1}{1-v^2}}-\frac{q^2}{2}\big(1-\frac{3}{\sqrt{1-v^2}}\big)\right)+\frac{v\sqrt{\lambda}a^2}{48\pi}\Big(\frac{1-v^2+\sqrt{1-v^2}+(1+v^2)\log(1+\sqrt{1-v^2})}{(1-v^2)^{3/2}}\nonumber\\
&&+\frac{q^2}{2(v^2-1)^2(1+\sqrt{1-v^2})}\big(v^4(-1-7\sqrt{1-v^2}-15(2-\sqrt{1-v^2})\log2)\nonumber\\
&&+v^2(2(-2+\sqrt{1-v^2})+15(4+\sqrt{1-v^2})\log2)-5(1+\sqrt{1-v^2})(-1+6\log2)\nonumber\\
&&-(v^2(1+v^2)(-2+\sqrt{1-v^2})+4(1+\sqrt{1-v^2}))\log(1+\sqrt{1-v^2})\big)\Big),\label{draganl}
\eea
where we have used $\ta=\pi/2$ and $\ta=0$, which correspond to $x-$ direction and $z-$ direction respectively. Note that the first part of right hand side of (\ref{drag_anl}) is
isotropic force in the charged plasma $F_{iso}$:
\bea
F_{iso}(q)=\frac{\pi\sqrt{\lambda}T^2v}{2}\left(\sqrt{\frac{1}{1-v^2}}-\frac{q^2}{2}\big(1-\frac{3}{\sqrt{1-v^2}}\big)\right),\label{F_iso}
\eea
and, when $q=0$, we can get the drag forces in the plasma for zero chemical potential, which coincide with the result of \cite{ma3}.
We can see from (\ref{F_iso}) that, in the isotropic case, the drag force in the charged plasma  is always larger than in neutral plasma at the same temperature.
In \cite{gi}, the author obtained a critical velocity $v_c \approx 0.909$ for anisotropic neutral case  independent of the temperature and anisotropy.
The critical velocity turns out to be charge--dependent and $v_c$ increases as $q$ increases.
\begin{table}[ht]
\begin{center}
\begin{tabular}{|c|c|c|c|c|c|}
         \hline
~$q$~ &~$0$~&~$0.04$~&~$0.06$~&~$0.08$~&~$0.1$~
          \\
        \hline
~$v_c$~ & ~$0.909$~ & ~$0.912$~ & ~$0.915$~&~$0.920$~&~$0.926$~
          \\
         \hline
\end{tabular}
\caption{\label{Tablev3}Critical velocity for different $q$.}
\end{center}
\end{table}
In the absence of the chemical potential, equations (\ref{draganl}) recover the analytical expressions for the transverse and parallel drag forces given in \cite{gi}.


\begin{thebibliography}{999}


\bibitem{rhic}
STAR Collaboration J.~Adams {\it et al.},
 Experimental and theoretical challenges in the search for the quark  gluon
plasma: The STAR collaboration's critical assessment of the  evidence from
RHIC collisions,
Nucl.\ Phys.\  A {\bf 757} (2005) 102 [nucl-ex/0501009].

\bibitem{rhic2}
PHENIX Collaboration ,K. Adcox {\it et al.},
 Formation of dense partonic matter in relativistic nucleus
collisions at RHIC: Experimental evaluation by the PHENIX  collaboration,
Nucl.\ Phys.\  A {\bf 757} (2005) 184 [nucl-ex/0410003].


\bibitem{duality}
  J. M. Maldacena,
   The Large N limit of superconformal field theories and supergravity,
  Adv.\ Theor.\ Math.\ Phys.\  {\bf 2} (1998) 231 [hep-th/9711200].

\bibitem{duality2}
  S.~S.~Gubser, I. R. Klebanov, A. M. Polyakov,
   Gauge theory correlators from noncritical string theory,
  Phys.\ Lett.\  {\bf B428} (1998) 105 [hep-th/9802109].

 \bibitem{duality3}
  E. Witten,
   Anti-de Sitter space and holography,
  Adv.\ Theor.\ Math.\ Phys.\  {\bf 2}  (1998) 253 [hep-th/9802150].

\bibitem{Hartnoll}
S.A.Hartnoll,
Lectures on holographic methods for condensed matter physics;
arXiv:0903.3246

\bibitem{McGreevy}
John McGreevy,
Holographic duality with a view toward many-body physics,
arXiv:0909.0518

\bibitem{Hartnoll:2008vx}
  S.~A.~Hartnoll, C.~P.~Herzog and G.~T.~Horowitz,
  ``Building a Holographic Superconductor,''
  Phys.\ Rev.\ Lett.\  {\bf 101}, 031601 (2008)
  [arXiv:0803.3295 [hep-th]].

\bibitem{Wen:2013ufa}
  W.~Y.~Wen, M.~S.~Wu and S.~Y.~Wu,
  ``A Holographic Model of Two-Band Superconductor,''
  Phys.\ Rev.\ D {\bf 89}, 066005 (2014)
  [arXiv:1309.0488 [hep-th]].

\bibitem{gw}
  X. H. Ge, B. Wang, S. F. Wu and G. H. Yang,
  ``Analytical study on holographic superconductors in external magnetic field ,''
  JHEP 1008, 108(2010)
  [ arXiv:1002.4901 [hep-th]].

\bibitem{Kachru:2008yh}
  S.~Kachru, X.~Liu and M.~Mulligan,
  ``Gravity duals of Lifshitz-like fixed points,''
  Phys.\ Rev.\ D {\bf 78}, 106005 (2008)
  [arXiv:0808.1725 [hep-th]].
\bibitem{Sun:2013wpa}
  J.~R.~Sun, S.~Y.~Wu and H.~Q.~Zhang,
  ``Novel Features of the Transport Coefficients in Lifshitz Black Branes,''
  Phys.\ Rev.\ D {\bf 87}, 086005 (2013)
  [arXiv:1302.5309 [hep-th]].

\bibitem{Sun:2013zga}
  J.~R.~Sun, S.~Y.~Wu and H.~Q.~Zhang,
  ``Mimic the optical conductivity in disordered solids via gauge/gravity duality,''
  Phys.\ Lett.\ B {\bf 729}, 177 (2014)
  [arXiv:1306.1517 [hep-th]].

  \bibitem{fang}
  L.~Q.~Fang, X. H. Ge and X. M. Kuang,
  `` Holographic fermions in charged Lifshitz theory,"
   Phys.Rev. {\bf D 86} 105037 (2012)
   [ arXiv:1201.3832 [hep-th]]



 \bibitem{review1}
  J.~Casalderrey-Solana, H. Liu, D. Mateos, K. Rajagopal, U.~A.~Wiedemann,
   Gauge/String Duality, Hot QCD and Heavy Ion Collisions,
  arXiv:1101.0618.

 \bibitem{review2}
 Youngman Kim, Ik Jae Shin, Takuya Tsukioka
 Holographic QCD: Past, Present, and Future,
 arXiv:1205.4852.

\bibitem{CaronHuot:2006te}
  S.~Caron-Huot, P.~Kovtun, G.~D.~Moore, A.~Starinets and L.~G.~Yaffe,
  ``Photon and dilepton production in supersymmetric Yang-Mills plasma,''
  JHEP {\bf 0612}, 015 (2006)
  [hep-th/0607237].

\bibitem{Wu:2013qja}
  S.~Y.~Wu and D.~L.~Yang,
  ``Holographic Photon Production with Magnetic Field in Anisotropic Plasmas,''
  JHEP {\bf 1308}, 032 (2013)
  [arXiv:1305.5509 [hep-th]].

\bibitem{Muller:2013ila}
  B.~Muller, S.~Y.~Wu and D.~L.~Yang,
  ``Elliptic flow from thermal photons with magnetic field in holography,''
  Phys.\ Rev.\ D {\bf 89}, no. 2, 026013 (2014)
  [arXiv:1308.6568 [hep-th]].

\bibitem{dts}
G. Policastro, D. T. Son and A. O. Starinets,  The shear viscosity of strongly coupled N
= 4 supersymmetric Yang-Mills plasma, Phys. Rev. Lett. 87 (2001) 081601 [hepth/
0104066].

\bibitem{drag1} C. P. Herzog, A. Karch, P. Kovtun, C. Kozcaz, L. G. Yaffe,
Energy loss of a heavy quark moving through N=4 supersymmetric Yang-Mills plasma,JHEP 0607 (2006) 013,
[hep-ph/0605158]

\bibitem{drag2}
Steven S. Gubser, Drag force in AdS/CFT,Phys. Rev. D 74, 126005,
[hep-ph/0605182]


\bibitem{jet1}
Sang-Jin Sin, Ismail Zahed, Holography of Radiation and Jet Quenching, Phys.Lett. B608 (2005) 265, [
hep-th/0407215].

\bibitem{jet2}
Hong Liu, Krishna Rajagopal, Urs Achim Wiedemann, Calculating the Jet Quenching Parameter from AdS/CFT, Phys. Rev. Lett. 97, 182301 (2006)
[hep-ph/0605178]

\bibitem{gil}
D. Giataganas  and H. Soltanpanahi, Universal Properties of the Langevin Diffusion Coefficients, Phys. Rev. D 89, 026011 (2014) [arXiv:1310.6725]


\bibitem{cuprates}
J. R. Schrieffer, James S. Brooks, Handbook of high temperature supercodncutvitity, Springer Press, 2007

\bibitem{Yee:2009vw}
  H.~U.~Yee,
  ``Holographic Chiral Magnetic Conductivity,''
  JHEP {\bf 0911}, 085 (2009)
  [arXiv:0908.4189 [hep-th]].

\bibitem{Kharzeev:2010gd}
  D.~E.~Kharzeev and H.~U.~Yee,
  ``Chiral Magnetic Wave,''
  Phys.\ Rev.\ D {\bf 83}, 085007 (2011)
  [arXiv:1012.6026 [hep-th]].

\bibitem{Pu:2014cwa}
  S.~Pu, S.~Y.~Wu and D.~L.~Yang,
  ``Holographic Chiral Electric Separation Effect,''
  Phys.\ Rev.\ D {\bf 89}, 085024 (2014)
  [arXiv:1401.6972 [hep-th]].

\bibitem{Pu:2014fva}
  S.~Pu, S.~Y.~Wu and D.~L.~Yang,
  ``Chiral Hall Effect and Chiral Electric Waves,''
  arXiv:1407.3168 [hep-th].

\bibitem{ch1}
L. Cheng,
X. H. Ge,  S.J. Sin, Anisotropic plasma with a chemical and schemen-idenpendent instablities, Phys. Lett. B 734 (2014) 116
[arXiv:1404.1994]


\bibitem{ch2}
 L. Cheng, X.H. Ge and S. J. Sin, Anisotropic plasma at finite $U(1)$ chemical potential, JHEP 07 (2014) 083 [arXiv:1404.5027]

 \bibitem{gl}
  X.H. Ge, Y. Ling, C. Niu and S. J. Sin,
Holographic transports and stability in anisotropic linear axion model, [arXiv:1412.8346]



\bibitem{ma1}
  D. Mateos and D. Trancanelli,  Thermodynamics and instabilities of a strongly coupled anisotropic plasma,  JHEP 1107 (2011) 054
  [arXiv:1106.1637].

\bibitem{ma2}
  D. Mateos and D. Trancanelli, The anisotropic N=4 super Yang-Mills plasma and its instabilities ,  Phys. Rev. Lett. 107 (2011) 101601
[arXiv:1105.3472].

\bibitem{rehban} A. Rebhan and D. Steineder,`` Violation of the holographic viscosity bound in a strongly coupled anisotropic plasma," Phys. Rev. Lett. 108 (2012) 021601 [arXiv:1110.6825]

\bibitem{ma3}
  M. Chernicoff, D. Fernandez, D. Mateos and Diego Trancanelli,
 Drag force in a strongly coupled anisotropic plasma,  JHEP 1208 (2012) 100 [arXiv:1202.3696].

\bibitem{gi}
  D.~Giataganas,
  Probing strongly coupled anisotropic plasma,
   JHEP {\bf 1207},(2012) 031
  [arXiv:1202.4436].

\bibitem{ma4}
M. Chernicoff, D. Fernandez, D. Mateos and Diego Trancanelli,
   Jet quenching in a strongly coupled anisotropic plasma, JHEP 1208 (2012) 041 [arXiv:1203.0561].

\bibitem{Chakraborty:2014kfa}
  S.~Chakraborty and N.~Haque,
  ``Drag force in strongly coupled, anisotropic plasma at finite chemical potential,''
  arXiv:1410.7040 [hep-th].

\end{thebibliography}
\end{document}